\documentstyle[astro,astrobib,rotating]{ar209}
\newcommand{\kms}{\mbox{km s$^{-1}$}}
\begin{document}
\def\altaffilmark#1{$^{#1}$}
\def\altaffiltext#1#2{\footnotetext[#1]{#2}\stepcounter{footnote}}
\def\eps@scaling{.95}
\def\epsscale#1{\gdef\eps@scaling{#1}}
\def\plotone#1{\centering \leavevmode
\epsfxsize=\eps@scaling\columnwidth \epsfbox{#1}}
\input epsf.sty   
\input psfig.sty  

\newcommand{\gb}{\gamma_{\rm j}\beta_{\rm j}}
\renewcommand{\topfraction}{0.85}
\renewcommand{\textfraction}{0.1}
\renewcommand{\floatpagefraction}{0.75}

\jname{Annu. Rev. Astron. Astrophys.}
\jyear{2001}
\jvol{39}
\ARinfo{ }

\title{The Supermassive Black Hole at the Galactic Center}

\markboth{Fulvio Melia and Heino Falcke}{The Galactic Center}
\altaffiltext{1}{Sir Thomas Lyle Fellow and Miegunyah Fellow.}

\author{Fulvio Melia\altaffilmark{1}
\affiliation{Physics Department and Steward Observatory, The University
of Arizona, Tucson, AZ 85721}\vskip -0.3in
Heino Falcke
\affiliation{Max-Planck-Institut f\"ur Radioastronomie, Auf dem H\"ugel 69, 
Bonn D-53121, Germany}}

\begin{abstract}
The inner few parsecs at the Galactic Center have come under intense 
scrutiny in recent years, in part due to the exciting broad-band 
observations of this region, but also because of the growing interest from 
theorists motivated to study the physics of black hole accretion, magnetized
gas dynamics and unusual star formation. The Galactic Center is now known
to contain arguably the most compelling supermassive black hole candidate, 
weighing in at a little over $2.6$ million suns. Its interaction with the nearby
environment, comprised of clusters of evolved and young stars, a molecular
dusty ring, ionized gas streamers, diffuse hot gas, and a hypernova
remnant, is providing a wealth of accretion phenomenology and high-energy 
processes for detailed modeling.  In this review, we summarize the latest 
observational results, and focus on the physical interpretation of the most 
intriguing object in this region---the compact radio source Sgr A*, thought
to be the radiative manifestation of the supermassive black hole. 

\medskip
\centerline{\it to appear in: Annual Review of Astronomy \&
Astrophysics, Vol. 39 (2001)}
\medskip

\end{abstract}

\maketitle

\section{INTRODUCTION}

The region bounded by the inner few parsecs at the Galactic Center 
contains six principal components that coexist within the central deep
gravitational potential well of the Milky Way.  These constituents are a
supermassive black hole, the surrounding cluster of evolved and young
stars, a molecular dusty ring, ionized gas streamers, diffuse hot gas,
and a powerful supernova-like remnant. Many of the observed phenomena
occurring in this complex and unique portion of the Galaxy can be
explained by the interaction of these components.

Though largely shrouded by the intervening gas and dust, the Galactic
Center is now actively being explored observationally at radio,
sub-millimeter, infrared, X-ray and $\gamma$-ray wavelengths with
unprecedented clarity and spectral resolution.  The interactions
governing the behavior and evolution of this nucleus are attracting
many astronomers and astrophysicists interested in learning about the
physics of black hole accretion, magnetized gas dynamics and unusual
stellar formation, among others.  The Galactic Center is one of the
most interesting regions for scientific investigation because it is
the closest available galactic nucleus and therefore can be studied
with a resolution that is impossible to achieve in other galaxies.
One arcsecond at the Galactic Center distance of $\sim 8$ kpc corresponds
to only $0.04$ pc ($\approx 1.2\times 10^{17}$ cm.
Thus, developing a consistent theoretical picture of the phenomena we
observe there improves not only our understanding of the Galaxy, but
also our view of galactic nuclei in general.

For example, the Galactic Center is now known to harbor by far the
most evident condensation of dark mass, which is apparently coincident
with the compact radio source Sgr~A*, the primary subject of this
review.  An overwhelming number of observations (proper and radial
motion of stars and gas) now strongly supports the idea that this
compact radio source in the center of the Galaxy has a mass of
$2.6\times 10^6\; M_\odot$ (see \S\ \ref{sgramass}). Because of these
unique observations and the proximity of Sgr~A*, the supermassive
black hole paradigm for galactic cores may be strengthened or refuted
based on what we learn about the Galactic Center.

The properties of Sgr~A* are, of course, not independent of its
environment. For example, one might naively expect from the observed
nearby gas dynamics, that Sgr~A* should be a bright source. Yet it is
underluminous at all wavelengths by many orders of magnitude,
radiating at only $3\times 10^{-10}$ of its Eddington luminosity. Does
this imply new accretion physics (as has been proposed) or does it
imply something peculiar about Sgr~A* itself? What now makes asking
these questions meaningful is that the extensive sets of data seriously
constrain the currently proposed answers.

Over the past decade the number of papers appearing in refereed
journals dealing with the theory of phenomena in the Galactic Center,
particularly the physics of Sgr~A*, has doubled roughly every three
years. The rate at which papers on the Galactic Center appear is now
more than one per week.  It is our intention here to summarize the
principal observational constraints, and to focus on the key
theoretical questions now facing the growing number of astrophysicists
working in this field.

\section{THE GALACTIC CENTER COMPONENTS}

\begin{figure}[thb]
\centerline{\psfig{figure=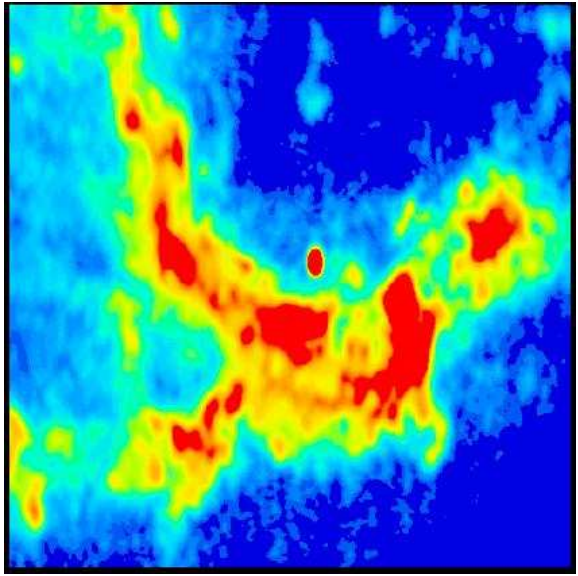,height=18pc}}
\caption{\scriptsize Sub-arcsec (2 cm) image of Sgr~A West and Sgr~A*.
The cometary-like feature to the north of Sgr A* (identified as the bright central
spot in this image) is associated with the luminous star IRS 7, seen at the corresponding
location in Fig. 2. (From Yusef-Zadeh \& Wardle 1993.)\label{fig-sgrawest}}
\end{figure}

\begin{figure}[thb]
\hskip 1.15in{\begin{turn}{90}
{\centerline{\psfig{figure=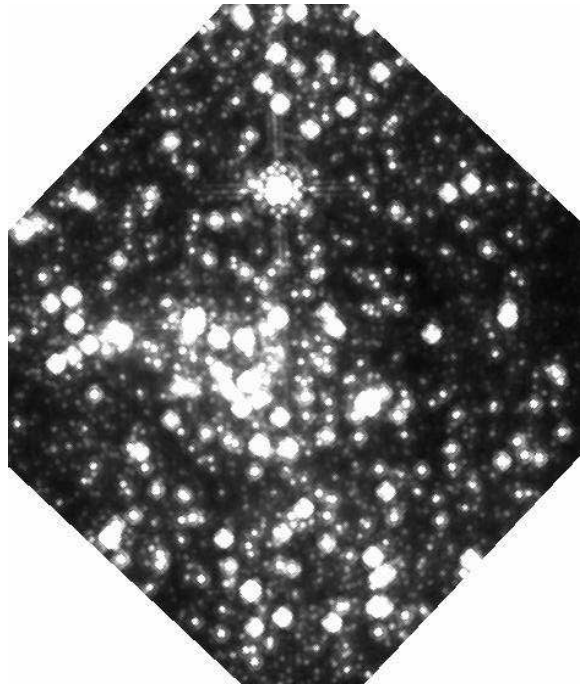,height=18pc}}}
\end{turn}}
\vskip -0.7in
\caption{\scriptsize NICMOS $1.6\mu$m image of the inner $19^{\prime\prime}\times
19^{\prime\prime}$ region at the Galactic Center. This is the same field
as that shown in Fig.~\ref{fig-sgrawest}. (Image courtesy of M. Rieke, Steward 
Observatory.)\label{fig-stars}}
\end{figure}

It is thought that the dynamical center of the Galaxy coincides
with Sgr~A$^*$ (\citeNP{EckartGenzelHofmann1995,MentenReidEckart1997,GhezKleinMorris1998}),
a compact non-thermal radio source no bigger than $\sim$
1~AU (see \S\ \ref{sgra}; \citeNP{KrichbaumZensusWitzel1993,BackerZensusKellermann1993,LoBackerKellermann1993,RogersDoelemanWright1994,KrichbaumGrahamWitzel1998,LoShenZhao1998}).  
On a slightly larger scale, the ``three-arm" spiral
configuration of ionized gas and dust known as Sgr~A West
\cite{EkersvanGorkomSchwarz1983,LoClaussen1983} engulfs this source in
projection (Fig.~\ref{fig-sgrawest}; here shown in a $\sim
2\;\hbox{\rm pc}\times 2$ pc image).  Figure~\ref{fig-stars} shows the
stellar distribution at $1.6\mu$m (on roughly the same spatial scale as
Fig.~\ref{fig-sgrawest}) as seen by NICMOS on the Hubble Space
Telescope {\it HST}. 
Sgr A* is in the very middle of this field of view, though it is not
seen at this wavelength.

\nocite{Yusef-ZadehWardle1993}

Sgr~A West probably derives its heat from the central distribution of
bright stars, rather than from a single point source (such as Sgr
A*; \citeNP{ZylkaMezgerWard-Thompson1995,Gezari1996,ChanMoseleyCasey1997,LatvakoskiStaceyGull1999}).
Some hot and luminous stars are thought to have been formed as
recently as a few million years
ago \cite{TamblynRieke1993,NajarroHillierKudritzki1994,KrabbeGenzelEckart1995,FigerKimMorris1999}.
We therefore see a sprinkling of several infrared-bright sources
throughout Sgr~A West that are probably embedded luminous stars, some
of which may be extended \cite{Gezari1996,TannerGhezMorris1999}.  It is
not yet clear whether these particular stars have formed within the
streamer or just happen to lie along the line of sight (see Fig.~\ref{fig-stars}).  

\vskip -0.6in

\begin{figure}[thb]
{\centerline{\psfig{figure=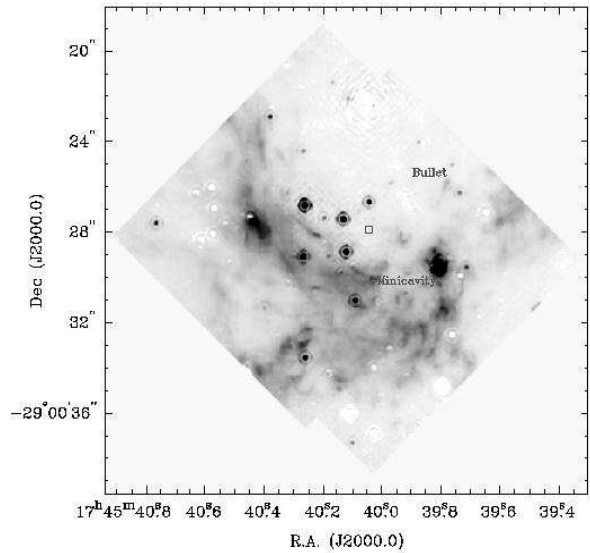,height=18pc}}}
\caption{\scriptsize Pa $\alpha$ mosaic of the central parsec (logarithmic inverse
gray scale with a dynamic range of 100) with $0\farcs2$ resolution.  Note
the wispy, filamentary structure in the ionized gas and the
sharp-edged bubble to the lower right known as the ``minicavity".
The fast-moving ``bullet" is also shown (the label is northeast of
the feature). A box centered on Sgr A* is shown for reference. Some
stars are also visible. (From Stolovy et al.~1999.)\label{fig-paa}}
\end{figure}

Spectroscopy of the hot gas in the mini-spiral structure seen
in Figure 1 
\cite{SerabynLacyTownes1988,HerbstBeckwithForrest1993,RobertsYusef-ZadehGoss1996}
suggests that it is rotating with a velocity of about 150 \kms\  around
Sgr~A* in a counter-clock wise direction; this confirms the inference
drawn from the Very Large Array (VLA) proper motion
studies \cite{Yusef-ZadehRobertsBiretta1998,ZhaoGoss1998,ZhaoGoss1999},
which have in addition shown the presence of high-velocity
features---such as the ``bullet" (see Fig.~\ref{fig-paa})---between 400 and 1,200 \kms. The
stream is evidently tugging along a milli-Gauss magnetic field seen through MIR
polarization imaging \cite{AitkenSmithGezari1991,AitkenSmithMoore1998} with
projected field lines aligned with the flow.

\nocite{StolovyMcCarthyMelia1999}

One of the most striking structures in Sgr~A West is the ``minicavity" centered
near the junction of its northern and eastern arms.  This feature,
adjacent to the peculiar sources Sgr~A$^*$ and IRS16 (the bright blue stellar 
cluster to the east of Sgr~A$^*$), is a distinct hole in the distribution of 
the radio continuum emission (Fig.~\ref{fig-sgrawest}) and the Pa $\alpha$ emission
(Fig.~\ref{fig-paa}) with a diameter of $2^{\prime\prime}$, corresponding 
to a linear dimension of 0.08 pc.  It may have been created by a spherical wind,
the source of which is yet to be identified, or it may be due in part
to the effects of a focused gas flow from the direction of Sgr A*
\cite{MeliaCokerYusef-Zadeh1996}.

\begin{figure}[thb]
\centerline{\psfig{figure=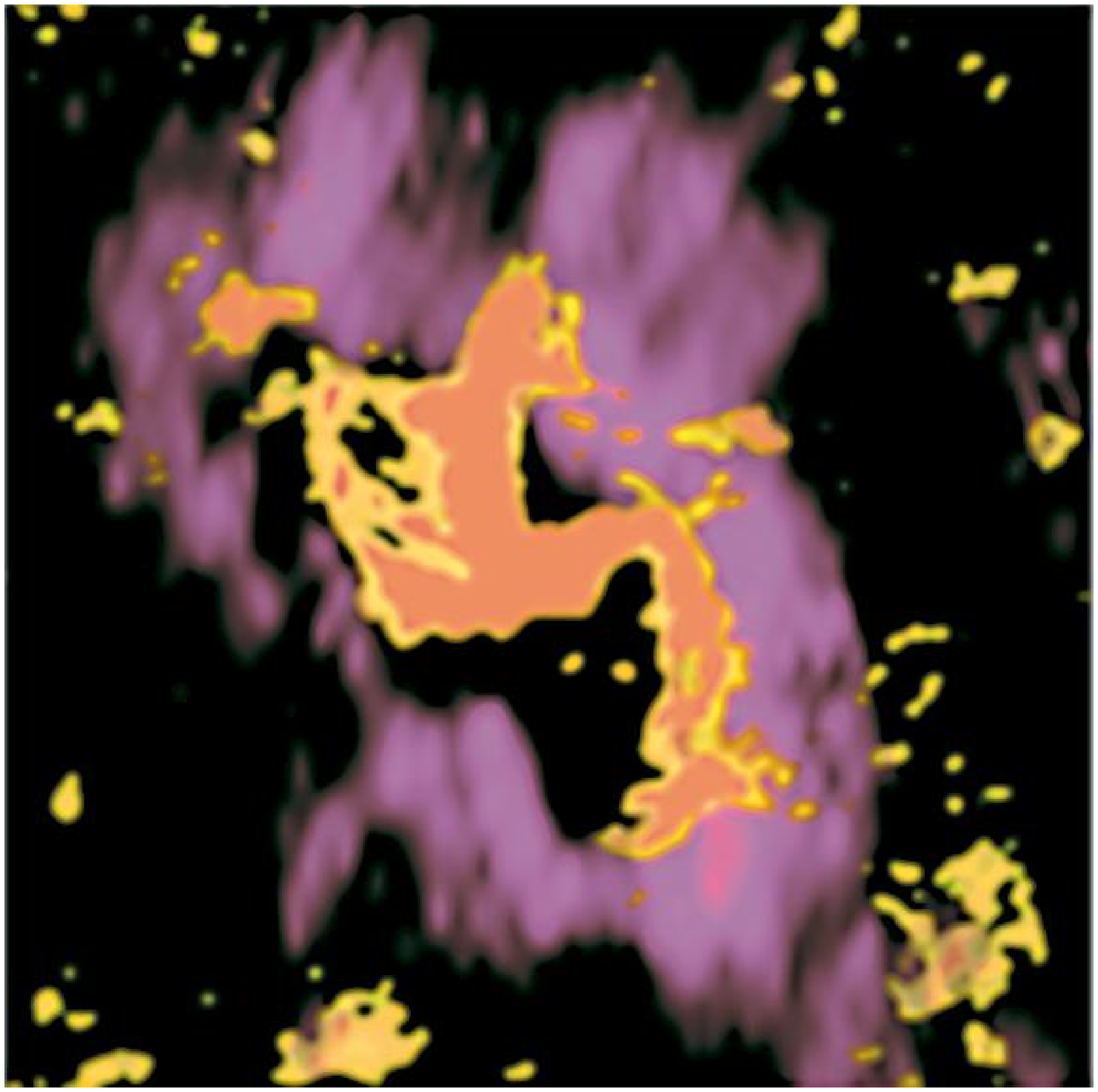,height=18pc}}
\caption{\scriptsize A radio image of ionized gas (Sgr~A West) at $\lambda=1.2$cm with
its three-arm appearance, shown in orange, superimposed on the distribution
of HCN emission, displayed in red (Wright et al.~1993).  Most of the ionized
gas is distributed in the molecular cavity.  At the distance to the Galactic
Center, this image corresponds to a size of approximately 4 pc on each side.
(From Yusef-Zadeh, Melia, \& Wardle 2000.)\label{fig-HCN}}
\end{figure}
\nocite{Yusef-ZadehMeliaWardle2000}

On an even larger scale ($\sim 3$ pc), Sgr~A West is thought to lie
within a large central cavity that is surrounded by a gaseous
circumnuclear ring (or circumnuclear disk, CND; 
\citeNP{BecklinGatleyWerner1982,DavidsonWernerWu1992,LatvakoskiStaceyGull1999,ZylkaGustenPhilipp1999})
and is otherwise relatively devoid of neutral gas, with the possible
exception of a ``tongue" of atomic gas that appears to be falling in
from the north \cite{JacksonGeisGenzel1993}.  A superposition of the 
radio continuum emission from Sgr~A West due to free-free radiation 
with an image showing the distribution of molecular gas 
(Fig.~\ref{fig-HCN}) suggests that this central cavity is filled with 
a bath of ultraviolet radiation heating the dust and gas within the 
inner 8 pc of the galaxy \cite{Yusef-ZadehStolovyBurton1999}.

\begin{figure}[thb]
\centerline{\psfig{figure=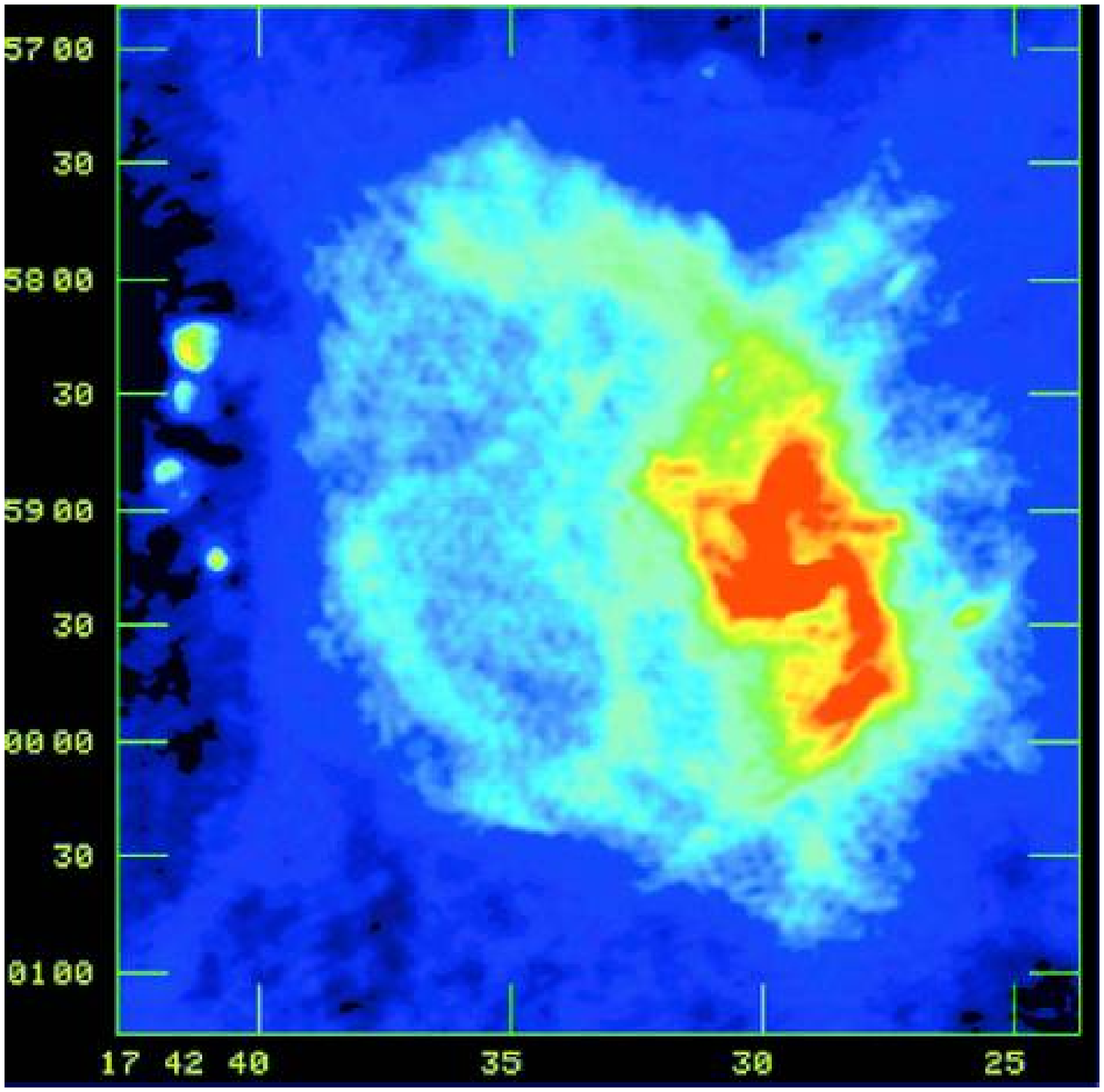,height=20pc}}
\caption{\scriptsize VLA radio continuum image of the Galactic
Center showing the shell-like structure of the non-thermal Sgr~A East
(light blue and green) and the spiral-shaped structure of the thermal
Sgr~A West (red) at $\lambda=6$cm with a resolution of $3.4\arcsec\times
2.9\arcsec$.  A cluster of HII regions associated with Sgr~A East
is also evident to the east of the shell.  The weak extended features
(dark blue) surrounding the shell are part of the Sgr~A East halo.
(From Yusef-Zadeh, Melia, \& Wardle 2000.)\label{sgraeast}}
\end{figure}

Radio continuum measurements of the larger-scale ($\sim 50\;\hbox{\rm pc}
\times 50$ pc) distribution of hot gas, known as the Sgr A complex, 
show a rather complicated morphology.
Recent improvements in spatial resolution and large scale imaging with
the VLA have helped considerably in separating the thermal and
non-thermal features in this region \cite{EkersvanGorkomSchwarz1983,Yusef-ZadehMorris1987,PedlarAnantharamaiahEkers1989}.
Sgr~A West and Sgr~A East constitute the brightest of the
continuum features (see Fig.~\ref{sgraeast}).  Sgr~A East could be a supernova
remnant (perhaps a bubble driven by several supernovae) or a very
low-luminosity example of a radio component associated with the active
nucleus of a spiral
galaxy \cite{PedlarAnantharamaiahEkers1989}. Observations of Sgr~A East
show it to be associated with the prominent 50 km s$^{-1}$ molecular
cloud near the Galactic Center.  Such an association would require more than
$10^{52}$ erg of explosive energy to account for the origin of Sgr~A
East \cite{MezgerZylkaSalter1989}, making it rather a hypernova 
remnant---perhaps due to a tidally disrupted star \cite{KhokhlovMelia1996}. 
On the other hand, recent X-ray observations \cite{MaedaBaganoffFeigelson2001} may
favor a classification as a young ($\sim 10^4$ yr) metal-rich mixed-morphology
supernova remnant.

The Sgr~A complex is also associated with diffuse X-ray 
emission \cite{PredehlTruemper1994,KoyamaMaedaSonobe1996,SidoliMereghetti1999,BaganoffBautzBrandt2001}.  The large temperature 
and pressure of the emitting region producing the hard X-rays suggest 
that this gas is probably unbound. The size of this feature, and 
its sound speed, argue for an age of $\sim$50,000 years for the hot 
plasma bubble \cite{KoyamaMaedaSonobe1996}.

On a scale of hundreds of parsecs, several
synchrotron-emitting filamentary structures run roughly in the
direction perpendicular to the Galactic plane \cite{Yusef-ZadehMorrisChance1984,Liszt1985,BallyYusef-Zadeh1989,GrayCramEkers1991,LangMorrisEchevarria1999,ReichSofueMatsuo2000}.
These filaments are possibly magnetic field lines (pervading the Galaxy)
that are lit up by relativistic electrons, e.g., in a reconnection
zone between the Galactic field lines and those from molecular clouds,
injected with ionized particles from a hot star
cluster \cite{SerabynMorris1994,FigerMorrisGeballe1999}.

The geometry of the Galactic magnetic field is generally thought to be poloidal
within $\sim100$ pc of the nucleus \cite{Morris1994,SofueLang1999} with a milli-Gauss 
intensity \cite{KilleenLoCrutcher1992,PlanteLoCrutcher1995,Roberts1999}.
The field lines also appear to be stretched in the azimuthal direction within
molecular clouds \cite{Novak1999,NovakDotsonDowell2000}.

The morphology of the large scale region at the Galactic Center is 
very rich.  More detailed accounts of these observations are provided 
in the reviews by \citeN{MorrisSerabyn1996} and 
\citeN{MezgerDuschlZylka1996}. A beautiful (and detailed) large-scale
view of the Galactic Center at 90 cm wavelength is presented by
\citeN{LaRosaKassimLazio2000}.  Below, we shall concentrate on the 
phenomenology and theory of the most enigmatic object within this 
array of sources at the Galactic Center---the supermassive black hole 
candidate, Sgr~A*.

\section{PHENOMENOLOGY OF SGR A*}\label{sgra}

\subsection{The Discovery of Sgr~A*}\label{sgrahistory}
The prescient application of the then very speculative black hole
model for quasars led \citeN{Lynden-BellRees1971} to point out that the
Galactic Center also should contain a supermassive black hole, perhaps
detectable with radio interferometry. Subsequently,
\citeN{BalickBrown1974} indeed found a compact radio source with the
National Radio Astronomy Observatory (NRAO) 
interferometer at Green Bank, later to be confirmed by
Westerbork \cite{EkersGossSchwarz1975} and Very Large Baseline
Interferometry (VLBI) observations
\cite{LoSchilizziCohen1975}. Eight years after its discovery, the
unresolved source was named Sgr~A* by \citeN{Brown1982} to distinguish
it from the more extended emission of the Sgr~A complex, and to
emphasize its uniqueness. More precise
high-resolution Very Large Array (VLA) observations
\cite{BrownJohnstonLo1981} indicated that it was located near the
dynamical center of the gas streamers in the Galactic nucleus, as
inferred from infrared fine-structure lines (\ion{Ne}{2};
\citeNP{LacyTownesGeballe1980}). Its radio variability was established
at about this time \cite{BrownLo1982}. The accumulation of these
observational signatures make it clear that Sgr~A* is a very unusual
object, rendering it a prime suspect for the location of the putative
supermassive black hole.

\newpage
\subsection{The Concentration of Dark Matter}\label{sgramass}

\vskip -1.1in
\begin{figure}[thb]
\hskip 0.5in{\begin{turn}{-90}
\centerline{\psfig{figure=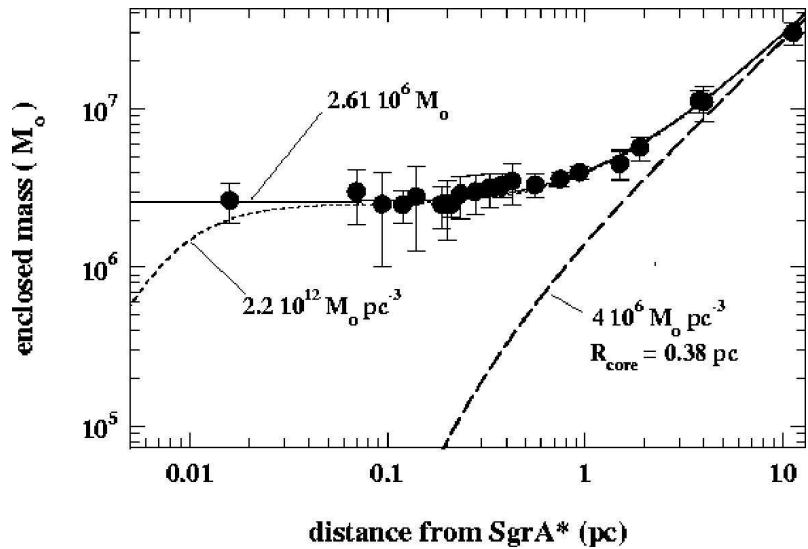,height=25pc}}
\end{turn}}
\vskip -0.9in
\caption{\scriptsize A plot of the distribution of enclosed mass versus distance
from Sgr~A*.  The three curves represent the mass model for a nearly
isothermal stellar cluster with a core radius of $0.38$ pc (thick
dashed line), the sum of this cluster plus a point mass of $2.61\pm
0.35\times 10^6\;M_\odot$ (thin solid curve), the same cluster and a
dark cluster with a central density of $2.2\times 10^{12}\;M_\odot$
pc$^{-3}$ and a core radius of $0.0065$ pc (thin dotted curve).  (From 
Genzel \& Eckart 1999.)\label{enclosedmass}}
\end{figure}
\nocite{GenzelEckart1999}

In their review, \citeN{GenzelTownes1987} published the now
well-known diagram showing the enclosed mass versus distance 
from Sgr~A*, suggesting a concentration of matter with
a point-like object (of mass $\sim3\times10^6M_{\sun}$) at the 
Galactic Center. This estimate depended rather sensitively on the 
mass inferred from the ionized gas motions 
\cite{SerabynLacy1985,SerabynLacyTownes1988}, which some thought
could have been influenced by non-gravitational forces (e.g., magnetic 
fields, stellar winds, etc.).  Even so, it was difficult to see how
the observed stellar winds and the measured magnetic fields in this general 
region could be strong enough to produce the observed velocities.  
In addition, infall from a large distance would have difficulty accounting
for the patterns seen \cite{Townes1996}.  The evidence for the existence of a dark
mass concentration has significantly and steadily grown since 
then---mainly via infra-red observations of stars near Sgr~A*.

The distribution of stellar radial velocities was inferred from
spectroscopic measurements, first for late-type giants and AGB stars
\cite{RiekeRieke1988,SellgrenMcGinnBecklin1990} 
and later for the hot ``\ion{He}{2}-stars"---blue supergiants 
close to their Wolf-Rayet stage \cite{NajarroKrabbeGenzel1997}---down 
to a distance of 1\arcsec~from Sgr~A*
\cite{KrabbeGenzelEckart1995,HallerRiekeRieke1996,GenzelThatteKrabbe1996}.
The most recent breakthrough has been provided by near-infrared
speckle imaging methods (i.e., shift-and-add techniques) that facilitated
the creation of a remarkable set of stellar proper motion data 
acquired over a six year-period with the ESO NTT, and later with Keck
\cite{EckartGenzel1996,EckartGenzel1997,GhezKleinMorris1998}. 
These measurements trace the stellar trajectories down to a scale as small 
as 5 light days from Sgr~A$^*$.

\begin{figure}[thb]
\centerline{\psfig{figure=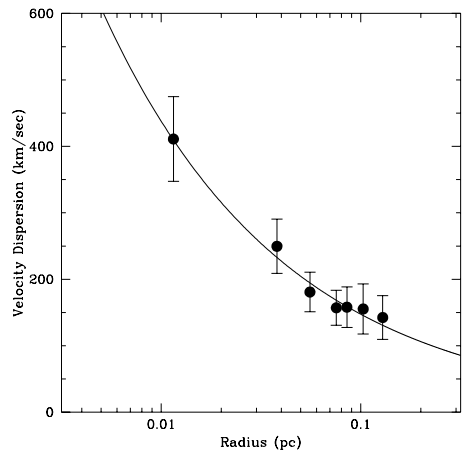,height=15pc}}
\caption{\scriptsize The projected stellar velocity dispersion
versus the distance from Sgr~A*.  The solid curve represents Keplerian
motion due to a mass concentrated within $0.01$ pc.  These data were obtained
with the Keck telescope.  (From Ghez et al.~1999.)\label{velocitydispersion}}
\end{figure}

The suggested central dark mass within the inner 0.015 pc of the
Galactic Center is $2.61\pm0.35\times 10^6\;M_\odot$.  ($0\farcs1$
corresponds to 800 Astronomical Units, or roughly $1.2\times10^{16}$ cm
at a distance of 8 kpc.) The inferred
distribution of matter as a function of distance from Sgr~A* is shown
in Figure \ref{enclosedmass}, and the measured stellar velocity
dispersion (shown in the accompanying Fig.~\ref{velocitydispersion})
is fully consistent with Keplerian motion about a highly compact
central mass concentration.  The value of these observations cannot be
overstated, since they establish the presence of a dark mass in the
Galactic Center beyond a reasonable doubt, even though several
systematic uncertainties (on a 10\% level) still remain; these include
the exact distance to the Galactic Center ($\sim8$ kpc; see
\citeNP{Reid1993}) and the exact mass estimator used to convert
velocities to masses.  The characteristic size associated with such a mass 
is the Schwarzschild radius $r_s\equiv 2GM/c^2$, which is here
equal to $7.7\times 10^{11}$ cm.  At a distance of $8$ kpc, this 
corresponds to $6.4\,\mu$as.

However, showing that the Galactic Center must contain a centralized
mass concentration does not yet necessarily imply that this dark
matter is in the form of an ultra-compact object with a few million solar
masses.  Nor does it exclusively imply that the unusual radio source
Sgr~A* must be associated with it; but it is possible to demonstrate
that Sgr~A* is not star-like, based on its position and proper
motion, which we consider next.

\nocite{GhezMorrisBecklin1999}

\subsection{Position and Proper Motion of Sgr~A*}
\begin{figure}[thb]
\centerline{\psfig{figure=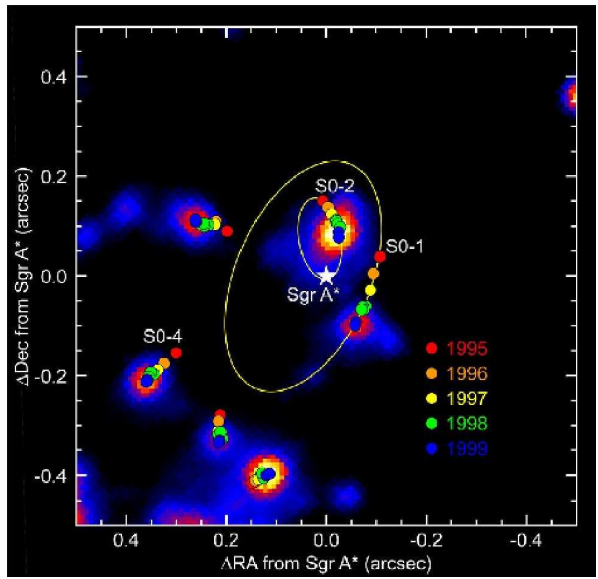,height=18pc}}
\caption{\scriptsize The orbits of two stars (labeled S0-1 and S0-2)
around Sgr A* as inferred from the detection of acceleration in their 
proper motion. The individual positions at the various epochs are shown 
as colored dots. The underlying image is a K-band Keck telescope image of 
the stars in the Galactic Center.  (From Ghez et al.~2000)\label{fig-orbits}}
\end{figure}

To begin with, how well does Sgr~A* actually coincide with the 
dynamical center of the stellar cluster? Thanks to the pioneering
work of \citeN{MentenReidEckart1997}, who found an SiO maser in the
bright star IRS 7 with the VLA, the location of Sgr~A* in the
near-infrared frame is now known to within 30 mas. Source counts
\cite{EckartGenzelHofmann1993} 
of the Near Infrared (NIR) stars show the center of the distribution
coincides with Sgr~A* to 
within fractions of an arcsecond. Similarly, using an unbiased
approach to identify stellar proper motions,
\citeN{GhezKleinMorris1998} find that the gravitational potential peaks 
on Sgr~A* within $\sim0\farcs1$. Recently, 
\citeN{GhezMorrisBecklin2000b} announced the detection of the first
signs of acceleration in the motion of stars allowing one to calculate
their orbits. The first results indicate that the dynamical center of
these bound trajectories coincides with Sgr~A* to within about $50$ mas
(Fig.~\ref{fig-orbits}). Another interesting by-product of these measurements 
is that the assumption of Keplerian motion may also help us to determine the 
distance to the Galactic Center more precisely \cite{SalimGould1999}.

Of equal significance is the argument first advanced by
\citeN{Backer1996}, that a heavy object in
dynamical equilibrium with the surrounding stellar cluster will move
slowly, so that a failure to detect random proper motion in Sgr~A* may be
used as a balance with which to weigh it.  In fact, such
measurements have been carried out using the VLA (for about 16 years;
\citeNP{BackerSramek1999a}) and the VLBA 
\cite{ReidReadheadVermeulen1999}, yielding consistent results of
a similar quality. For example, the latter yield a proper motion in Sgr
A* of $-3.33\pm0.1$ E and $-4.94 \pm 0.4$ N mas yr$^{-1 }$,
which corresponds to $-5.90 ±\pm 0.35$ and
$+0.20 \pm 0.30$ mas yr$^{-1}$ in Galactic longitude and latitude,
respectively. This apparent motion 
amounts to a constant velocity that is entirely consistent with
the 220 km s$^{-1}$ rotation of our solar system around the Galactic
Center. The position of Sgr~A* at 1996.25 in J2000 coordinates was
\begin{equation}
{\rm RA}(1996.25) = 17^{\rm h}45^{\rm
m}40\fs0409, \quad {\rm DEC}(1996.25) = -29\degr00\arcmin28\farcs118\;,
\end{equation} 
with an absolute uncertainty of 12 mas \cite{ReidReadheadVermeulen1999}. After
removal of the Galactic rotation, the upper limit to any proper motion
intrinsic to Sgr~A* is about $\pm15$ km s$^{-1}$.  This implies that 
our basic understanding of Galactic structure seems to be correct and 
that Sgr~A* is indeed located in the center of the Galaxy. 

These observations also provide a lower bound to Sgr A*'s mass. 
Clearly, most stellar objects in this region have
transverse velocities that statistically should peak around 
100 - 200 km s$^{-1}$---an order of magnitude greater than the upper limit
for Sgr~A*; in the vicinity of this object, stellar motions
reach 1,000 km s$^{-1}$, or more. Assuming no central point mass and an
equipartition of the momentum \cite{ReidReadheadVermeulen1999} 
between the fastest stars ($m_* v_*$) and
Sgr~A* $(M_{\rm Sgr\,{A^*}}\,v_{\rm Sgr\,{A^*}})$ one naively infers a 
mass
\begin{equation}
M_{\rm Sgr\,A*}\ga1,000\, M_{\sun}
\left({m_*\over10\,M_{\sun}}\right)\left({v_*\over1,500\,{\rm
km/s}}\right)\left({v_{\rm Sgr\,A*}\over15\,{\rm km/s}}\right)^{-1}\;,
\end{equation}
arguing for a non-stellar nature of Sgr~A*.  This rules out any possible
identification with a pulsar or a neutron star. 

Simple N-body simulations by
\citeN{ReidReadheadVermeulen1999} show that the momentum exchange
between the stars and Sgr~A* during close encounters probably
offers the dominant contribution to the latter's proper motion, which
for a $2.6\times10^6M_\odot$ black hole is less than 0.1 km
s$^{-1}$, i.e., consistent with the observations. On the other hand, if
Sgr~A* did not mark the location of the dark mass, it would certainly feel
its potential. To avoid seeing any motion in Sgr~A* one would
have to conclude either that its orbit around a compact mass is
extremely small ($\ll1$ mas)---thus requiring essentially a point mass
again---or that the mass distribution is rising extremely steeply (i.e.,
a Plummer model with $\alpha=5$) between the VLBI scale ($\sim1$ mas,
4$\times10^{-5}$ pc) and the stellar motion scale (300 mas, $\sim$0.01
pc). Again, even in this very contrived case the N-body simulations require a
strict lower limit of 1,000 $M_{\sun}$ for Sgr~A*.

The NIR stellar proper motion studies and the radio positional
measurements of Sgr A* are complementary.  Together, they constitute
a compelling argument in favor of Sgr A* defining the dynamical
center of the central star cluster, and therefore of the Galaxy. 
In this respect, it might be worth considering shifting the origin
of the Galactic coordinate system ($l=0,b=0$) to the location of Sgr A*.
\subsection{Size Constraints and the Brightness Temperature}\label{sgrsize}
\begin{figure}[thb]
\hskip -0.4in\vbox{
\centerline{\psfig{figure=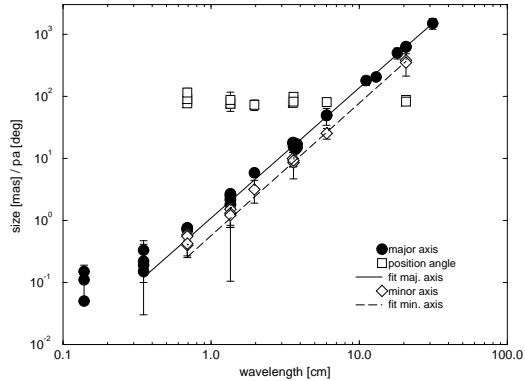,height=0.25\textheight,angle=-90}}}
\caption{\scriptsize The major source axis (filled circles) 
of Sgr\,A*, the minor source axis (open diamonds)
and the position angle of the major axis (open squares) as measured by 
VLBI plotted versus wavelength.  (From Krichbaum et al.~1999.)\label{scattersize}}
\end{figure}

The main problem in determining the size of Sgr~A* is that its true
structure is washed out by scattering in the interstellar medium 
\cite{DaviesWalshBooth1976,vanLangeveldeFrailCordes1992,Yusef-ZadehCottonWardle1994,LoShenZhao1998},
leading to a $\lambda^2$ dependence of its diameter as a function of
the observed wavelength (Fig.~\ref{scattersize}). 
Some of the underlying theory is discussed in
\citeN{RomaniNarayanBlandford1986}. The scattering is anisotropic,
possibly because of large scale magnetic fields pervading the inner
Galaxy
\cite{Yusef-ZadehCottonWardle1994}, with a roughly constant 
ratio between the major and minor axes of 0.53 at all frequencies below 43
GHz and a constant position angle of $80\pm3^\circ$. The functional
form of the scattering size is given by
\citeN{LoShenZhao1998} as
\begin{equation}
\theta_{\rm minor}=0.76\,{\rm mas}\;(\lambda/{\rm cm})^2\quad\theta_{\rm
major}=1.42\,{\rm mas}\;(\lambda/{\rm cm})^2\;,
\end{equation}
and the scattering size apparently has not changed over a decade
\cite{MarcaideAlberdiLara1999}.

\nocite{KrichbaumWitzelZensus1999}

However, it is possible to constrain the mm-to-sub-mm {\it intrinsic} size of Sgr~A* 
to within a factor of 10. From the absence of refractive scintillation,
\citeN{GwinnDanenTran1991} have argued that Sgr~A* must be larger than
$10^{12}$ cm at $\lambda\lambda1.3$ and $0.8$mm.
An upper limit to its size comes from VLBI observations at
mm-wavelengths. Because of the low elevation of Sgr~A* in most of
these observations, the NS resolution is usually much poorer than the
EW resolution. The problem of interpreting an elongated source
structure in Sgr~A* with insufficient baseline coverage is discussed
in \citeN{DoelemanRogersBacker1999} and
\citeN{BowerFalckeBacker1999b}. The most recent measurements were
carried out by \citeN{BowerBacker1998}, who find a source size of
$9\times10^{13}$ cm at 43 GHz ($\lambda7$mm)---a mere 2 $\sigma$
above the scattering size. Significantly, \citeN{LoShenZhao1998} infer an elongated
source in the North-South direction with a size of $5.5\times(1.5\times
10^{13}$ cm$)$. Together, these results indicate a 4 $\sigma$ deviation from the
scattering size and, perhaps unexpectedly, appears in the minor axis size
dependence. Finally, observations at 86 GHz ($\lambda3$mm) and 215 GHz
($\lambda1.4$mm)
\cite{RogersDoelemanWright1994,KrichbaumGrahamWitzel1998,DoelemanShenRogers2001} 
demonstrate that Sgr~A* is compact on a scale at or below $0.1$ mas 
($1.3\times10^{13}$ cm) for the highest frequencies. This
corresponds to $\sim17$ Schwarzschild radii for a
$2.6\times10^6\;M_\odot$ black hole. While the exact size of Sgr~A*
cannot yet be stated with absolute certainty, the latest observations
fuel hopes that somewhere in the millimeter wave regime the intrinsic
source size will finally dominate over interstellar broadening, 
allowing a direct comparison with the predictions of various emission theories
(see \S\ ~\ref{sgraemission}).

The upper limit to Sgr~A*'s size ($\sim 1$ A.U.) requires that its
brightness temperature be greater than $\sim 10^{10}$ K.  Its minimum
size of $\sim 0.1$ A.U. corresponds to an
upper limit on the brightness temperature (at $0.8$ mm) of about 
$0.5\times 10^{12}$ K.  Sgr~A* therefore is within the range of typical
AGN radio cores \cite{Readhead1994} and shines below the Compton limit
(at approximately $10^{12}$ K; \citeNP{KellermannPauliny-Toth1969}).
This is the maximum brightness temperature of incoherent synchrotron
emission from an electron plasma.  Above this temperature, the
radiation is heavily Comptonized to a frequency well beyond the GHz
range.  However, if the emitting particles are Maxwellian, they must
reach relativistic energies, i.e., their temperature must exceed $\sim
5\times 10^9$ K, in order for them to be efficient synchrotron
emitters (see, e.g., \citeNP{Melia1992a,Melia1994,MahadevanNarayanYi1996}),
corresponding to electron Lorentz factors of a few and possibly
hundreds.

\subsection{The Spectrum of Sgr~A*}\label{submmbump}
After intense radio observation of this source over many years, the spectrum 
of Sgr~A* at these wavelengths is rather well known.  Unfortunately, several 
claims of counterpart identification at shorter wavelengths (NIR, MIR, X-rays) 
have turned out to be chance associations with other sources 
\cite{EckartGenzelKrabbe1992,RosaZinneckerMoneti1992,GoldwurmCordierPaul1994,StolovyHaywardHerter1996}. 

\citeN{DuschlLesch1994} compiled an average radio spectrum from the published
data and claimed a rough $\nu^{1/3}$ power-law. However, in simultaneous
multi-frequency VLA observations (e.g.,
\citeNP{WrightBacker1993,MorrisSerabyn1996,FalckeGossMatsuo1998}), 
the actual spectrum is seen to be bumpy and the spectral index at GHz
frequencies varies between $\alpha=0.1-0.4$ ($S_{\nu}\propto\nu^\alpha$) 
\cite{BrownLo1982,Falcke1999a}. There may be a
low-frequency turnover of the spectrum around 1 GHz
\cite{DaviesWalshBooth1976}, the nature of which has never been clarified in 
detail, though several suggestions (i.e., due to a scattering size that is 
too large, free-free absorption, and self-absorption) have
been proposed. At very low frequencies (e.g., 330 MHz) the entire
Sgr A region suffers from free-free absorption
\cite{PedlarAnantharamaiahEkers1989}.

At high frequencies, Sgr A*'s spectrum must also drop off steeply
due to its faintness in the infrared, which is somewhat out of
character for an AGN,
as first noted by \citeN{RiekeLebofsky1982}. One of the most
interesting features currently under study is the suggestion of a
sub-millimeter bump in the spectrum
\cite{ZylkaMezgerLesch1992,ZylkaMezgerWard-Thompson1995,SerabynCarlstromLay1997} since
in all emission models the highest frequencies correspond to the
smallest spatial scales, so that the sub-millimeter emission is
almost certainly coming directly from the vicinity of the
black hole \cite{Melia1992a,Melia1994}.

\begin{figure}[thb]
\centerline{\psfig{figure=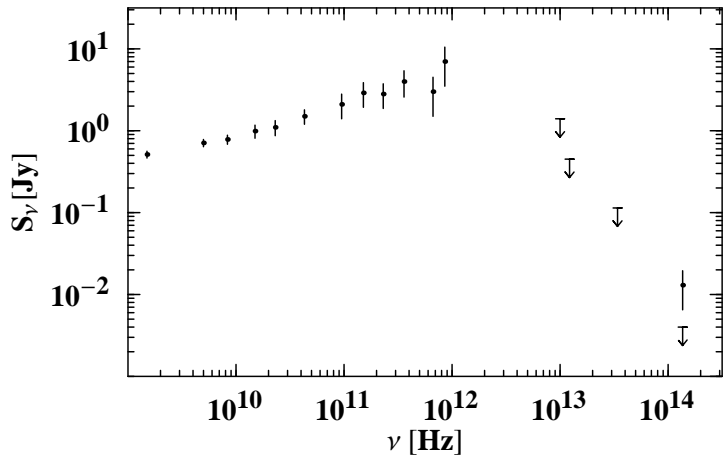,height=15pc,bbllx=4.1cm,bburx=16cm,bblly=19.2cm,bbury=27.1cm}}
\caption{\scriptsize Time-averaged spectrum---flux density versus frequency---of
Sgr A* from radio to the near infrared.  The radio data up to 22 GHz are
from Zhao et al. (2001) averaged over 1980-2000;  43 \& 95 GHz data
are the averages from the campaign of Falcke et al. (1998);
230 \& 350 GHz data are from Zylka et al. (1995) averaged over 1987-1994,
and 1991-1994, respectively;  the remaining data are discussed in
the text.  The error bars indicate variability (one standard deviation),
assumed to be at least $50\%$ beyond 350 GHz.\label{sgrspec}}
\end{figure}

However, the existence of this bump has been uncertain due to the
variability of Sgr~A*. In 1996, the spectrum of Sgr~A* was measured
simultaneously from $\lambda$20 cm to $\lambda$1 mm, with four different
telescopes (VLA, BIMA, Nobeyama 45 m, \& IRAM 30 m), on three
continents.  The results of this campaign are incorporated into the
averaged data plotted in Figure \ref{sgrspec},
which shows Sgr~A*'s spectrum ranging from $1.36$ to $232$ GHz. 

\nocite{ZhaoBowerGoss2001,ZylkaMezgerWard-Thompson1995,FalckeGossMatsuo1998}

The spectrum at lower frequencies was adequately described by two
power-laws with spectral indices $\alpha=0.17$
($S_\nu\propto\nu^\alpha$) between 1.36 GHz and 8.5 GHz and
$\alpha=0.30$ between 15 and 43 GHz.  The 20 year average spectral
index of Sgr A* in the range $1.4-22$ GHz is $\alpha=0.28$, as
derived by \citeN{ZhaoBowerGoss2001}.  The spectral index of Sgr~A*
increases to $\alpha=0.52$ in the mm-range, while the $\lambda$3-to-2
mm spectral index becomes even higher, reaching $\alpha=0.76$.  Based
on these results, one can safely conclude that there probably exists a
significant mm-excess in the spectrum of Sgr~A*, or possibly even a
separate component at these frequencies. So far no bright
confusing source has been found in high-resolution millimeter-wave
maps that could explain such an excess as being extrinsic to Sgr~A*
(e.g., \citeNP{ZhaoGoss1998}). However, it is possible that the excess,
seen also as a curvature in time-averaged spectra, is in part due to
large-amplitude flares at high frequencies \cite{TsuboiMiyazakiTsutsumi1999}.

In the sub-mm range, detections exist up to 666 GHz
\cite{ZylkaMezgerWard-Thompson1995} with a flux density of up to $3-4$ Jy.
\citeN{SerabynCarlstromLay1997} claim a $7\pm2$ Jy detection at 850 GHz
while only upper limits exist in the mid-infrared: $<1.4$ Jy at 30
$\mu$m \cite{TelescoDavidsonWerner1996}, $<450\pm150$ mJy at 24.5$\mu$m
and $<114\pm30$ mJy at 8.81 $\mu$m
\cite{CoteraMorrisGhez1999}\footnote{These flux densities were
dereddened by a factor $10^{\left({A_\lambda/ A_{\rm
V}}\right)\times A_{\rm V}\times 0.4}\sim5.7$ for an A$_{\rm V}=30$, using the
extinction law given in \citeN{RiekeLebofsky1985} and \citeN{Mathis1990},
with ${A_{8.81\mu{\rm m}}/ A_{\rm V}}=0.063$ and ${A_{24.5\mu{\rm m}}/ 
A_{\rm V}}=0.014$}. In the near-infrared, \citeN{GenzelEckart1999} 
give an upper limit of 4 mJy at 2.2 $\mu$m, with an (as yet unconfirmed) 
detection around 13 mJy of Sgr A* in certain epochs (all flux densities are 
dereddened).  In the optical, of course, Sgr ~A* is reddened by an 
A$_{\rm V}=30$ and hence undetectable.  A time-averaged spectrum from
$1.4$ GHz to $10^{14}$ Hz is shown in Figure 10.

In the X-rays, {\it ROSAT} detected emission at the position of Sgr~A*
\cite{PredehlTruemper1994}, but with a rather large beam that might have included some
diffuse emission as well. The luminosity in the 0.2-2 keV range was
about $1-2\times10^{34}$ erg s$^{-1}$ for standard Galactic reddening
\cite{PredehlZinnecker1996}. Recently {\it Chandra} detected a source within 
$1\arcsec$ of Sgr~A* with a rather low X-ray luminosity of about 
$0.9\times10^{34}$ erg s$^{-1}$ in the 0.5-10 keV band for a photon
index of $2.75^{+1.25}_{-1.0}$ \cite{BaganoffBautzBrandt2001}. This
appears to be the first convincing detection of Sgr~A* in an
energy range different from radio. There is also $\gamma$-ray
emission at a level of $(2.2\pm0.2)\times10^{37}$ erg s$^{-1}$ above
100 MeV \cite{Mayer-Hasselwander1998} from the Sgr A region.  This
level should be considered an upper limit for Sgr~A* because of 
the possibly very large extent ($\sim1.5^\circ$) of the emitter
and the lack of detected variability. Finally, there is also a claimed
excess of high-energy neutrons ($\sim 10^{18}$ eV) seen in air showers
toward the direction of the Galactic Center \cite{HayashidaNaganoNishikawa1999},
which may, however, also come from a relatively large region.

\subsection{Radio Variability of Sgr~A*}\label{variability}
An important parameter for constraining the spectrum and nature of Sgr
A* is its variability. In the radio flux density, variations are
clearly seen between different epochs, but the time scale of the
variability at various frequencies is not well determined and it is
not clear whether some of the more extreme claims of variability are
real or instrumental artifacts. \citeN{ZhaoEkersGoss1989} and
\citeN{ZhaoGossLo1992} found a number of outbursts at higher
frequencies and a rather low level of variability at low frequencies.

\citeN{Falcke1999a} published the results of 540 daily observations of 
Sgr~A* at 2.3 and 8.3 GHz with the Green Bank Interferometer (GBI).  A
peak-to-peak variability of 250 mJy with an RMS (i.e., modulation
index) of 6\% and 2.5\% at 8.3 and 2.3 GHz respectively was found. The
median spectral index between the two observed frequencies (8.3 and
2.3 GHz) for the whole period was $\alpha=0.28$
($S_\nu\propto\nu^\alpha$), varying between 0.2 and 0.4. There is a
clear trend for the spectral index to become larger when the flux
density in both bands increases. The spectral index versus flux
correlation and the different modulation indices at the two
frequencies imply that outbursts in Sgr~A* are more pronounced at
higher frequencies. This is not consistent with a simple model of
refractive interstellar scintillation as suggested by
\citeN{ZhaoGossLo1992}: the variability time scale inferred at 2.3 GHz
and 8.3 GHz is comparable to that found at 5 GHz by
\citeN{ZhaoGossLo1992} and it does not follow a $t\propto\lambda^2$
law.

The most direct conclusion one can draw from the variability data is
the high degree of correlation between emission at 2.3 and 8.3 GHz. The lag is
apparently less than three days which corresponds to a light travel
distance of $\leq10^{16}$ cm ($\sim60$ mas at the distance
to the Galactic Center;  this is less than the scattering size). For 
models that have a frequency-dependent structure (e.g., the accretion
and jet models) this will be an upper limit to the size of the emitting
region at these frequencies. At both frequencies the characteristic 
time scale is somewhere between 50 and 250 days. There is very little 
variability on time scales of a few days
below 10 GHz and a slow, linear increase of the flux density is
observed over the entire 2 years of observations.
 
Recently \citeN{ZhaoBowerGoss2001} have
investigated 20 years of VLA data and find marked outbursts with an
amplitude around 0.4 Jy at 23 GHz with a characteristic time scale of less than
25 days that are not seen below 8 GHz. The flare amplitude seems to
increase with frequency.  Similar flares at mm-waves have
also been observed by \citeN{TsuboiMiyazakiTsutsumi1999} and
\citeN{WrightBacker1993} even though mm-wave flux densities are notoriously
difficult to calibrate. An intriguing result that needs further
confirmation is the possibility that these high-frequency flares are
periodic (or perhaps quasi-periodic) with a period around 106 days
\cite{ZhaoBowerGoss2001}.  A theoretical interpretation for these
variability characteristics is provided in \S\ 4 below.

\subsection{The Measured Linear Polarization of Sgr~A*}
Early papers that discussed the radio emission of Sgr~A did not report
any significant polarization. \citeN{EkersGossSchwarz1975} quoted an
upper limit of 1\% linear polarization for the region of peak emission
in Sgr~A, which at that time was not well resolved in their 5 GHz
Westerbork observations.  Subsequent observations with the VLA
similarly yielded a null polarization measurement
\cite{Yusef-ZadehMorris1987}. This was in contrast to the situation
with AGNs, in which the linear polarization is typically a few percent
\cite{HughesAllerAller1985,MarscherGear1985}.  So, while the measurement 
of polarization promises useful information, the early negative results 
for Sgr~A* made this a non-issue for over a decade.  

One reason why the linear polarization in Sgr~A* is low could be the
presence of a scattering screen between the Galactic Center and the
observer, which de-polarizes the radiation. This can come about
because (1) the differential Faraday rotation of the homogeneous medium may be so
high that within the bandwidth of the observation the polarization
vector is rotated by more than 180 degrees and therefore is largely
canceling itself (see below), and (2) there may be considerable variation of the
Faraday rotation in the scattering screen so that every ray that
reaches the observer gets rotated differently, which reduces the
overall polarization significantly. In this context, it may be
relevant to ask whether such an effect could lead to a conversion from
linear to circular polarization; after all, we have an anisotropic
scattering screen permeated by a large scale magnetic field
\cite{Yusef-ZadehCottonWardle1994}.

\citeN{BowerBackerZhao1999} have reported the results of continuum polarimetry 
at 4.8 GHz and spectro-polarimetry at 4.8 and 8.4 GHz, using the VLA.
The spectro-polarimetric observations were made to exclude strong
differential Faraday rotation in the Galactic Center that could lead to a
de-polarization of the radiation when observed in continuum mode
integrating over a large bandwidth. Faraday rotation is produced when
radio waves pass through an ionized and magnetized medium. Since left
and right circularly polarized waves have different refractive indices
for a given magnetic field orientation, a wavelength-dependent delay
is induced which rotates the position angle $\phi_{\rm LP}$ of the
linear polarization vector, yielding $\phi_{\rm LP}={\rm
RM}\lambda^2$. The parameter RM is called the rotation measure and can
be determined by measuring the position angle of the linear
polarization vector $\phi_{\rm LP}$ at different wavelengths
$\lambda$. For a given frequency bandwidth $\Delta \nu$, significant
de-polarization is obtained if $\phi_{\rm LP}$ changes by more than
one radian, i.e., if ${\rm RM}>0.5 {\nu/(\lambda^2\Delta\nu)}$.  This
means that for a typical VLA bandwidth of 50 MHz at 4.8 GHz, the
critical rotation measure is $\sim10^4$ rad m$^{-2}$, which is not
deemed to be so excessively high that it could not be present in the
Galactic Center.

\begin{figure}
\centerline{\psfig{figure=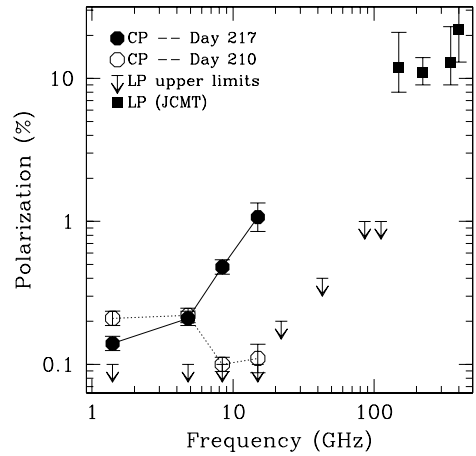,width=0.48\textwidth}}
\caption[]{\label{sgrpol}\scriptsize
Linear and circular polarization in Sgr~A* from 1.4 to 86 GHz. 
The down arrows indicate upper limits for linear polarization
measurements. The open octagons are CP measurements from the 
VLA on July 28, 1999. The filled octagons are measurements from
the VLA on August 5, 1999. The sign of CP has been flipped in this 
figure. Also shown is the 10+9-4\% detection of LP at 150 GHz by
Aitken et al.~(2000). (From Bower 2000.)}
\end{figure}
\nocite{Bower2000}

The results of the broad-band continuum polarimetry indeed confirmed
the absence of linear polarization with a rather low upper limit of
$<0.1\%$ fractional polarization.  By Fourier transforming the
spectro-polarimetric data to sample multiple rotations of the
polarization vector across the entire band, \citeN{BowerBackerZhao1999}
were able to exclude Faraday rotation as the cause of this, with RM
values up to $10^7$ rad m$^{-2}$.  To further clarify whether RM
fluctuations in the scattering medium could de-polarize the radiation
from Sgr~A*, one could simply try to measure the linear polarization
at progressively higher frequencies. Since the scattering size
decreases with $\nu^{-2}$ the differential changes in the angles to
the line of sight for light rays from Sgr~A* will rapidly become
smaller and smaller with increasing wavelength. In addition the
Faraday rotation itself will also decrease with
$\nu^{-2}$. \citeN{BowerWrightBacker1999} and
\citeN{BowerWrightFalcke2000} have sought high-frequency polarization
of Sgr~A* with the VLA and found only upper limits.  So
de-polarization of the radiation from Sgr~A* by the scattering medium
appears to be rather unlikely at present.
 
The situation at sub-mm wavelengths could be different.  The most
recent information is provided by linear polarization measurements
using the SCUBA camera at the James Clerk Maxwell Telescope (JCMT), at
0.75, 0.85, 1.35, and 2 mm \cite{AitkenGreavesChrysostomou2000}.
These authors have reported the detection of fractional linear polarization as
high as 10\% at these wavelengths. However, one potential problem with these
low-resolution observations is the possible confusion of the Sgr~A*
flux with that from dust emission in the surrounding circumnuclear
disk and from the mini-spiral in Sgr A West.  
If confirmed, the lack of a detected polarization at 
$\lambda$3.5 mm and $\lambda$2.7 mm, in contrast to these detections 
at shorter wavelengths, may be a possible signature of compact sub-mm
emission from within several Schwarzschild radii of the black hole 
\cite{AitkenGreavesChrysostomou2000,QuataertGruzinov2000b,Agol2000,MeliaLiuCoker2000}.

\subsection{The Measured Circular Polarization of Sgr~A*}
In synchrotron sources, the degree of circular polarization is $m_c <
0.1\%$; only rarely has $m_c$ reportedly reached $0.5\%$
\cite{WeilerdePater1983}.  The degree of circular polarization usually
peaks near 1.4 GHz and decreases strongly with increasing frequency.

Given the stringent upper limits on the linear polarization of
Sgr~A* at cm-wavelengths, the detection of circular polarization
came as somewhat of a surprise \cite{BowerFalckeBacker1999}.
The circular polarization in Sgr~A* was found to be $m_c=-0.36 
\pm 0.05\%$ and $m_c=-0.26 \pm 0.06\%$ at 4.8 and 8.4 GHz. 
The detection was quickly confirmed by \citeN{SaultMacquart1999},
using the Australia Telescope Compact Array at 4.8 GHz.

The circular polarization is variable on a ten day time scale and is now
detected up to 43 GHz \cite{BowerFalckeSault2000,Bower2000}. The
overall spectrum of circular polarization seems to be inverted and
increases beyond 8 GHz (see Fig.~\ref{sgrpol}).

Polarimetric measurements of Sgr~A* are opening up a relatively new and 
exciting field. The first attempts at interpreting the 
polarization properties of this source are just being made
\cite{Agol2000,QuataertGruzinov2000b,MeliaLiuCoker2000,BeckertFalcke2001} and 
some of these are discussed below in \S\ \ref{sgraemission}.

\section{GAS DYNAMICS AND STELLAR WIND CAPTURE}\label{gasdyn}
Having described the key observational characteristics of Sgr A*,
let us now turn our attention to the physical interpretation of this object. 
As we alluded to in the previous section, the abundance of gas in the
environment surrounding Sgr~A* clearly points to accretion as the
incipient cause of its ensuing energetic behavior. The
properties described above are consistent with the idea that
Sgr~A*'s spectrum results from the energy liberated by a compressed
hot plasma either bound to the central gravitational potential during
infall (see \S\ 5.1), or in the process of expulsion in the form 
of a jet (see \S\ 5.2).  However, we shall first play the ``devil's
advocate" and consider the possibility that the potential well is 
instead associated with a distributed cluster of dark objects
(rather than a single point mass; \citeNP{MeliaCoker1999}), and then 
compare the results with the expectations for a black hole potential.

\begin{figure}[thb]
\centerline{\psfig{figure=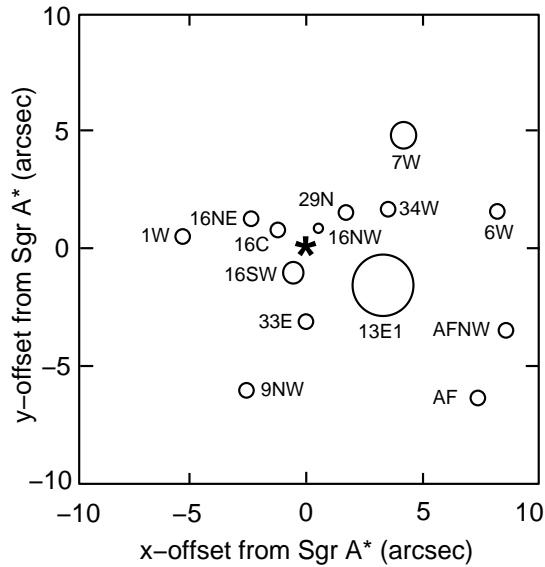,height=18pc}}
\caption{\scriptsize Location of some of the wind-producing stars relative to the
position of Sgr~A* indicated by the * symbol. The radius of each circle
corresponds (on a linear scale) to that star's mass loss rate.  Setting
the scale is 13E1, with $\dot M=7.9\times 10^{-4}\;M_\odot$ yr$^{-1}$. (From 
Coker \& Melia 1997.)\label{figwinds}}
\end{figure}

The Galactic Center wind appears to be produced by the early-type
stars enclosed (in projection) within the Western Arc, the Northern
Arm, and the Bar.  Thus far, 25 such stars have been identified
\cite{GenzelThatteKrabbe1996}, and all appear to be located within the central
parsec surrounding Sgr~A*.  Figure~\ref{figwinds} shows the positions
(relative to Sgr~A*) of these wind sources, in which the size of the
circle marking each position corresponds to the relative mass loss
rate (on a linear scale) for that star.  Note that due to clustering
some of the stars are combined into a single wind source in this
figure, and some are outside the field of view.

\begin{figure}[thb]
\centerline{\psfig{figure=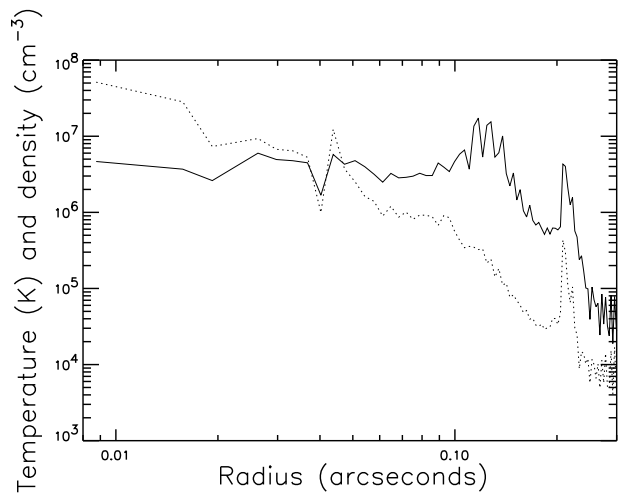,height=16pc}}
\caption{\scriptsize Angle- and volume-averaged density (dotted) and temperature
(solid) as a function of radius (in arcseconds) from the center. The
flattening of these distributions at small radii (i.e., $< 0.^{\prime\prime}1
\approx 0.004$ pc) is clearly evident.  (From Melia \& Coker 1999.)\label{winddensity}}
\end{figure}

The gravitational potential of a dark cluster can be represented by an
``$\eta$-model" \cite{HallerMelia1996}.  This function mimics an
isotropic mass distribution with a single parameter, and it is scaled
so that the total dark cluster mass within $0.01$ pc is $2.6\times10^6\;M_\odot$.
\citeN{MeliaCoker1999} used the 3D hydrodynamics code ZEUS to simulate
the flow of the Galactic Center wind through this distributed dark
matter using the $\eta$-potential, and one of the key results of this
calculation is summarized in Figure~\ref{winddensity}, which shows the
angle- and volume-averaged density and temperature for the whole
central $0.^{\prime\prime}7$ region, using a bin size of
$0.^{\prime\prime}0025$.  In this figure, the density rises gradually
to the middle, and reaches an average value of roughly $10^8$
cm$^{-3}$.  In contrast, the central density for a gas falling freely
into a black hole potential with the same central mass approaches
$\sim 10^{13}$ cm$^{-3}$ \cite{Melia1994}.  The temperature similarly
rises to the middle, but it levels off within about $0.004$ pc, and
the average is never greater than about $10^7$ K.  This is to be
compared with the temperature profile of the gas falling into the
black hole, where $T$ attains values as high as $10^{10}$ K or more.
This is critical because the electrons begin to emit significantly via
the synchrotron process when they become relativistic above a few
times $10^9$ K.  This gas can at best therefore only emit cyclotron
radiation, but the emissivity is a strong function of $T$ and is here
insignificant compared to bremsstrahlung.  The flattening of the
density and temperature profiles shown in Figure \ref{winddensity} is
a direct consequence of the shallowness of the cluster potential
compared to the steep potential gradients encountered by the gas
falling into the black hole.  The magnetic field, which is coupled to
the physical state of the gas, behaves in a similar fashion, though it
is clumpier due to the uneven dissipation in regions of gas
compression and rarefaction.

Gas flowing through a dark cluster may get trapped, but it clearly
does not produce the type of condensation and high temperature required 
to account for Sgr~A*'s spectrum.  This is indirect support for the inference
drawn from other lines of evidence that the dark matter is instead
concentrated in the form of a single compact object.

Let us therefore consider the physical state of the gas when the
gravitational potential well deepens as the plasma approaches the
event horizon. In the classical Bondi-Hoyle (BH) scenario
\cite{BondiHoyle1944}, the mass accretion rate for a uniform
hypersonic flow is $\dot M_{BH} = \pi {R_A}^2 m_H n_w v_w$, in terms
of the accretion radius $R_A \equiv 2 G M / {v_w}^2$.  With the
conditions at the Galactic Center (see above), we would therefore
expect an accretion rate $\dot M_{BH} \sim 10^{21}$ g s$^{-1}$ 
($\approx 1.6\times 10^{-5}\;M_\odot/$yr) onto
the black hole, with a capture radius $R_A \sim .02$ pc.

In reality the flow past Sgr~A* is not likely to be uniform, since
one might expect many shocks to form as a result of wind-wind
collisions within the IRS 16 complex, even before the plasma reaches
$R_A$.  The implications for the spectral
characteristics of Sgr~A*, and thus its nature, are significant.
\citeN{CokerMelia1997} have therefore undertaken the task of
simulating the BH accretion from the spherical winds of a distribution
of 10 individual point sources located at an average distance of a few
$R_A$ from the central object. The results of these simulations show
that the accretion rate depends not only on the distance of the
mass-losing star cluster from the accretor but also on the relative
spatial distribution of the sources.  In addition, the co-existence
of hot and warm gas components may itself alter the Bondi-Hoyle capture
profile \cite{BaganoffBautzBrandt2001}, which is not included in these
simulations.  The capture rate inferred by these authors is
$\approx 3\times 10^{-6}\;M_\odot/$yr.

\begin{figure}[thb]
\centerline{\psfig{figure=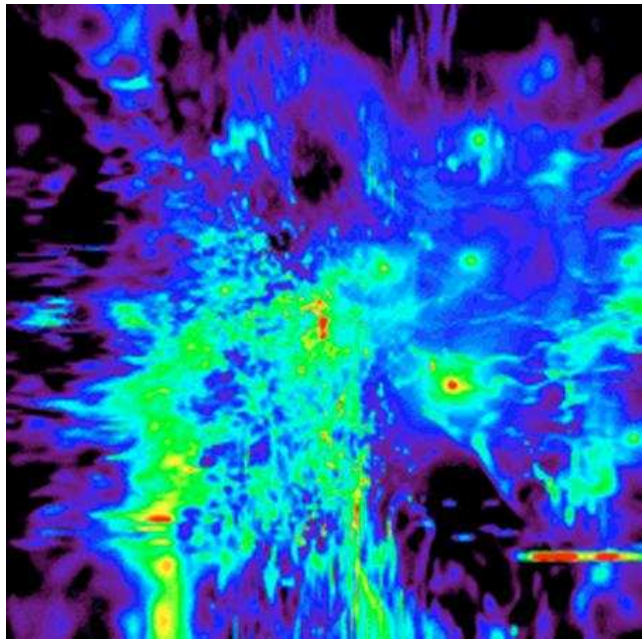,height=20pc}}
\caption{\scriptsize A ``snapshot" of the column density
(i.e., the gas density integrated along the line of sight) taken at a point in the calculation
when the gas distribution had reached stationary equilibrium.  Sgr~A* is in the middle, and
the dimensions are approximately $0.5$ light years on each side.  Some 15 to 20 stars surrounding
the black hole each produce an efflux of gas (i.e., ``winds"), which collide and form this
tessellated pattern of gas condensations, some of which are captured by the black hole and
accrete towards it.  Several of the wind-producing stars are visible to the right of the
image.  The color scale is logarithmic, with red corresponding to a column density
of $10^{21}$ g cm$^{-2}$, then yellow, blue, and black, which corresponds
to $10^{16}$ g cm$^{-2}$. (From Coker \& Melia 1997.)\label{figwinddistrib}}
\end{figure}

Figure~\ref{figwinddistrib} shows a logarithmic color scale image of
the density profile for a slice running through the center of the
accretor, for one of these simulations taken 2,000 years after the
winds are ``turned on".  Once the stellar winds have cleared the
region of the original low density gas, all such simulations point to
an overall average density ($\sim10^3$ cm$^{-3}$) in agreement with
observations.

\begin{figure}[thb]
\centerline{\psfig{figure=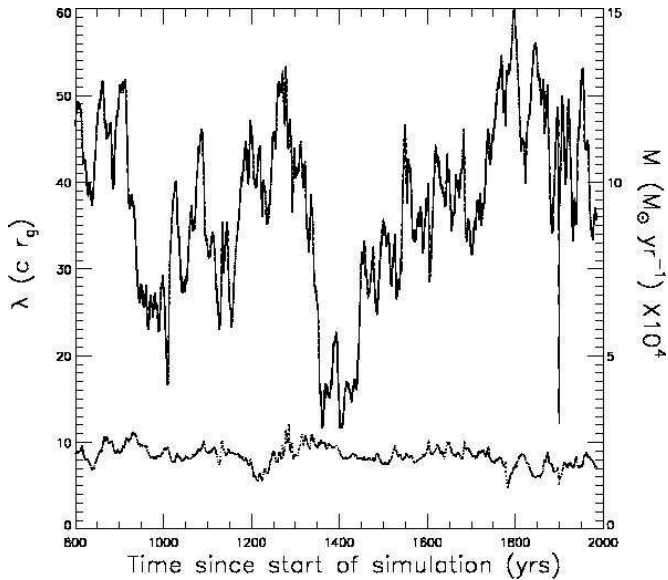,height=18pc}}
\caption{\scriptsize The upper solid curve is the {\it magnitude} of the
accreted specific angular momentum $\lambda$ (in units of $cr_s$).  
The scale for $\lambda$ is on the left side.
The lower dotted curve is the mass accretion rate $\dot M$ (${10}^{-4}\;M_\odot$ yr$^{-1}$)
versus time.  The scale for $\dot M$ is shown on the right side.  
(From Melia \& Coker 1999.)\label{figwindaccretion}}
\end{figure}

Figure~\ref{figwindaccretion} shows the mass accretion rate,
$\dot M$, and the accreted specific angular momentum, $\lambda$ (in
units of $c r_s$, where $r_s$ is the Schwarzschild radius), versus
time, starting 2 crossing times ($\sim$ 800 years) after the winds are
``turned on".  The average value for the mass accretion rate once the
system has reached equilibrium is $\dot M = 2.1 \pm 0.3 \dot M_{BH}$.

The mass accretion rate shows high frequency temporal fluctuations
(with a period of ${\lower.5ex\hbox{$\; \buildrel < \over \sim \;$}}
0.25$ yr) due to the finite numerical resolution of the simulations.
The low frequency aperiodic variations (on the order of $20\%$ in
amplitude) reflect the time dependent nature of the flow.  Thus, the
mass accretion rate onto the central object, and consequently the
emission arising from within the accretor boundary, is expected to
vary by ${\lower.5ex\hbox{$\; \buildrel < \over \sim \;$}} 20-40\%$
(since in some models the luminosity may vary by as much as $\propto
{\dot M}^2$) over the corresponding time scale of $<100$ years, even
though the mass flux from the stellar sources remains constant.  The
temporal variations in Sgr~A*'s radio luminosity
(\S\ \ref{variability}) are probably due (at least in part) to these
fluctuations in the accretion rate toward small radii.

Similarly, the accreted $\lambda$ can vary by $50\%$ over
${\lower.5ex\hbox{$\; \buildrel < \over \sim \;$}}$ 200 years with an
average equilibrium value of $37 \pm 10$.  It appears
that even with a large amount of angular momentum present in the wind,
relatively little specific angular momentum is accreted.  This is
understandable since clumps of gas with a high specific angular
momentum do not penetrate to within 1 $R_A$.  The variability in the
sign of the components of $\lambda$ suggests that if an accretion disk
forms at all, it dissolves, and reforms (perhaps) with a different 
sense of spin on a time scale of $\sim 100$ years or less.

The captured gas is highly ionized and magnetized, so it radiates via
brems\-strahlung, cyclo-synchrotron and inverse Compton processes.
However, the efficiency of converting gravitational energy into
radiation is quite small (as little as $10^{-4}$ in some cases), so
most of the dissipated energy is carried inwards
\cite{Shapiro1973,IpserPrice1982,Melia1992a,Melia1994}.  In fact, if
the magnetic field is a negligible fraction of its equipartition value
(see below), Sgr~A* would be undetectable at any frequency, except
perhaps at soft X-ray energies. But as the plasma continues to
compress and fall toward smaller radii, one or more additional things
can happen, each of which corresponds to a different theoretical
assumption, and therefore a potentially different interpretation,
which we explore in the next section.

\section{EMISSION MODELS FOR SGR A*}\label{sgraemission}
\subsection{Emission due to the Accreting Plasma}
The questions one may ask include the following: (1) Does the flow
carry a large specific angular momentum (in contrast to our
expectations from the Bondi-Hoyle simulations) so that it forms a disk
with lots of additional dissipation?  (2) Does the flow produce a radiatively
dominant non-thermal particle distribution at small radii (e.g.,
from shock acceleration), or does thermal emission continue to
dominate the spectrum?  (3) Does the flow lead to an expulsion of plasma
at small radii that forms a non-thermal jet, which itself may then
dominate the spectrum?  These, either individually or in combination,
have led to a variance of assumptions about the nature of the inflowing
gas that then form the basis for the development of different interpretations.

Observationally, one of the key issues is why the infalling gas maintains a low
radiative efficiency.  In the picture developed by \citeN{NarayanYiMahadevan1995},
and updated in \citeN{NarayanMahadevanGrindlay1998}, the infalling gas is 
assumed to carry a very large angular momentum towards the center, forming a big 
accretion disk (with an outer edge extending beyond $10^5$ Schwarzschild radii
or so). The Bondi-Hoyle simulations discussed above suggest that clumps of gas 
with relatively large specific angular momentum do not penetrate inwards.  
However, a large disk may form if the viscosity is anomolously high even
at large radii.  In this case, the overall emission must now include the 
additional dissipation of the captured angular momentum. To comply with the 
observed low efficiency of Sgr A*, this model therefore also assumes 
that the electron temperature is much lower than that of the protons ($T_e\ll
T_p$).  In fact, $T_e<10^{10}$ K.  Since the electrons do the radiating,
the efficiency remains small even though the protons are very hot.
It is important to point out, in this regard, that the success or
otherwise of an advection-dominated model rests on whether or not
event horizons really do exist.  The low efficiency of such an inflow
can be maintained only if the energy transported inward vanishes from
view \citeN{NarayanMahadevanGrindlay1998}. 

Large accretion disks such as this are known as ADAFs.  Strictly
speaking, the acronym ADAF stands for Advection Dominated Accretion Flow,
which embraces all forms of accretion (disk or otherwise) in which 
a large fraction of the dissipated energy is advected inwards by the
hot protons, rather than radiated away locally by the electrons.  So for
example, if the gas flow is quasi-spherical until it gets to within
a handful of Schwarzschild radii (as suggested by the Bondi-Hoyle
simulations) it may still be advection-dominated if the emissivity of the
gas is very low; this may occur when the magnetic field is weak 
(\citeNP{KowalenkoMelia1999,CokerMelia2000}; see below).  In practice, 
however, the term ADAF is conventionally used to denote the category of accretion 
patterns that involve a large, two-temperature {\it disk}.

Not surprisingly, the radiative and dynamic properties of ADAFs are sensitive 
to the outer boundary conditions, which are not well known. In their analysis, 
\citeN{YuanPengLuWang2000} adopted $T_e$, $T_p$, and the specific angular 
momentum of the accretion flow at the outer boundary as their principal 
free parameters.  Allowing these variables to range over reasonable 
values produces differences of several orders of magnitude in the peak 
radio, IR, and X-ray fluxes.  An additional complication is the possible 
``contamination" of the thermal particle distribution with non-thermal 
particles produced, e.g., from the decay of charged mesons, which are
themselves created through proton-proton collisions 
\cite{MarkoffMeliaSarcevic1997,MarkoffMeliaSarcevic1999,Mahadevan1999}.
Nonetheless, ADAF models can be designed to give reasonable fits to
the data \cite{MenouQuataertNarayan1999}, though the recent Chandra
X-ray measurements seem problematic (see below).

An important evolution in the ADAF theory came with the realization that
when a black hole accretes gas conservatively at a rate well below the Eddington
value (so that its radiative efficiency is very low), the net enery flux,
including the energy transported by the viscous torque, is likely to be
close to zero at all radii (\citeNP{BlandfordBegelman1999}; see also
\citeNP{NarayanYi1994}).  In other words, a 
large fraction of the plasma in an ADIOS (i.e., an Advection Dominated
Inflow/Outflow Solution) may be unbound, leading to significant mass
loss in the form of a wind.  As such, the assumption of a constant
accretion rate throughout the ADAF region may be quite poor.  This
situation is not unrelated to the Bondi-Hoyle result 
(\citeNP{CokerMelia1997}; see previous section) that clumps of gas with
large specific angular momentum generally do not accrete inwards.  
Much of the recent effort in this area has therefore been channelled
into producing more detailed, numerical simulations
to gauge whether the ADAF idea still remains viable as an explanation
for Sgr A*'s radiative characteristics.  Several independent groups
\cite{Hujeirat1999,ManmotoKatoNakamurNarayan2000,TurollaDullemond2000}
who are now investigating the structure of ADAF disks are reporting
positive Bernoulli values for a wide range of conditions, indicating 
that outflows are a necessity, though perhaps not as large as the 
early ADIOS estimates seemed to suggest;  for example, 
\citeN{TurollaDullemond2000} report a ratio of inflowing to 
outflowing mass of about $1/2$. Earlier,
\citeN{IgumenshchevChenAbramowicz1996} had shown that serious outflows
occur only if the viscosity parameter $\alpha$ is $0.3$ or larger,
which may be unrealistic, leaving somewhat uncertain the issue of
whether real outflows occur or not.

As more and more physical details are added to this study, the degree 
of complexity in the flow grows in corresponding fashion.  It now
appears that ADAF disks are also convectively unstable for low values
of viscosity \cite{IgumenshchevAbramowiczNarayan2000}.  Hydrodynamic
simulations of such flows reveal a radial density profile that is
significantly flatter than that expected for a canonical ADAF.  Other
recent modifications to the canonical ADAF model include the introduction 
of ADAFs without turbulent viscosity driving the accretion process.  In 
one such picture \cite{KinoKaburakiYamazaki2000}, accretion through the 
ADAF disk is instead controlled by a large-scale magnetic field.
Another new ingredient is the influence of convection (``CDAF", standing
for convective ADAF model), which requires an extremely low accretion
rate, around $10^{-8}\;M_\odot$ yr$^{-1}$ 
\cite{QuataertGruzinov2000a,QuataertGruzinov2000b,NarayanIgumenshchevAbramowicz2000}.
 
A difficulty faced by the ADAF disk model is that there does not appear to 
be a simple way out of the large dissipation (and consequent radiative 
efficiency) produced by the wind falling onto the plane \cite{FalckeMelia1997}. 
In addition, the ADAF model is yet to be established observationally.
ADAF disk models have now been applied extensively to several low-luminosity
systems, including the cores of elliptical galaxies, but compelling
observational evidence for their existence is lacking.  
\citeN{DiMatteoFabianRees1999} examined the high-frequency radio
observations of NGC 4649, NGC 4472, and NGC4636 and concluded that
the new radio limits disagree with the canonical ADAF
predictions, which tend to significantly overestimate the observed flux.
They concluded that if accretion in these objects occurs in an
advection-dominated disk mode, the radio limits imply a strong suppression
of the emission from the central regions. This problem may be worse still
since the measurements reported in this paper apparently included substantial 
extended emission (e.g., from the jet) due to the poor spatial resolution of 
the observations. 

A possible resolution to this problem is that the magnetic field within
the inflowing gas may be sub-equipartition, which clearly has the effect
of lowering the synchrotron emissivity.  This effect may be present whether 
or not the dissipated energy in the flow is advected inwards through the event 
horizon.  The idea that Sgr~A*'s low emissivity is due to a sub-equipartition 
magnetic field $B$ deserves close attention, 
especially in view of the fact that the actual value of $B$ depends strongly 
on the mechanism of field line annihilation, which is poorly understood.  Two 
processes that have been proposed are (i) the Petschek mechanism 
(\citeNP{Petschek1998}), in which dissipation of the sheared magnetic 
field occurs in the form of shock waves surrounding special neutral points in 
the current sheets and thus, nearly all the dissipated magnetic energy is 
converted into the magnetic energy carried by the emergent shocks; and (ii) the 
tearing mode instability \cite{vanHovenHendrixSchnack1995}, which relies on 
resistive diffusion of the magnetic field and is very sensitive to the physical
state of the gas.  In either case, the magnetic field dissipation rate is a 
strong function of the gas temperature and density, so that assuming a fixed 
ratio of the magnetic field to its equipartition value may not be appropriate.

\begin{figure}[thb]
\centerline{\psfig{figure=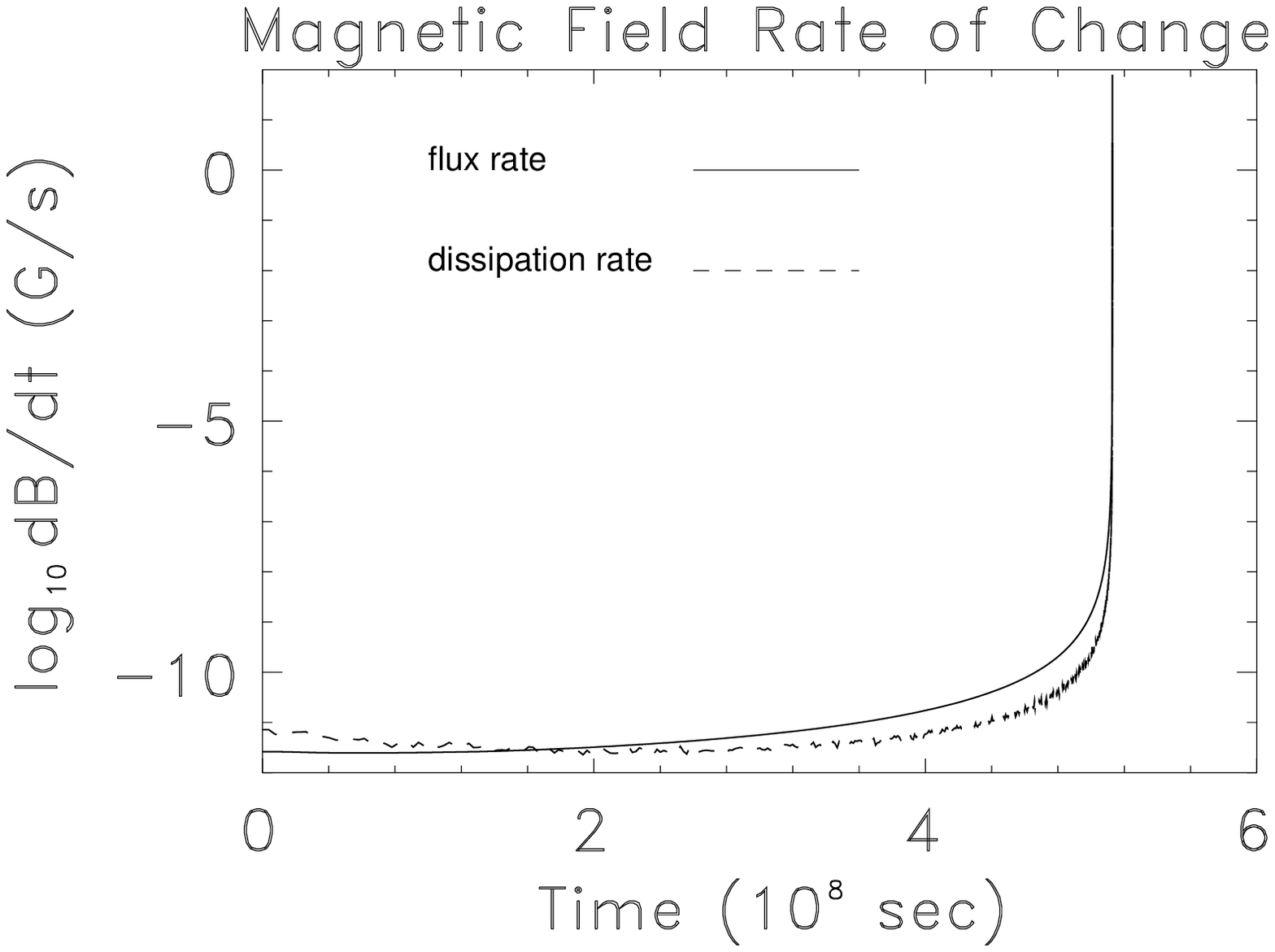,height=12pc}\psfig{figure=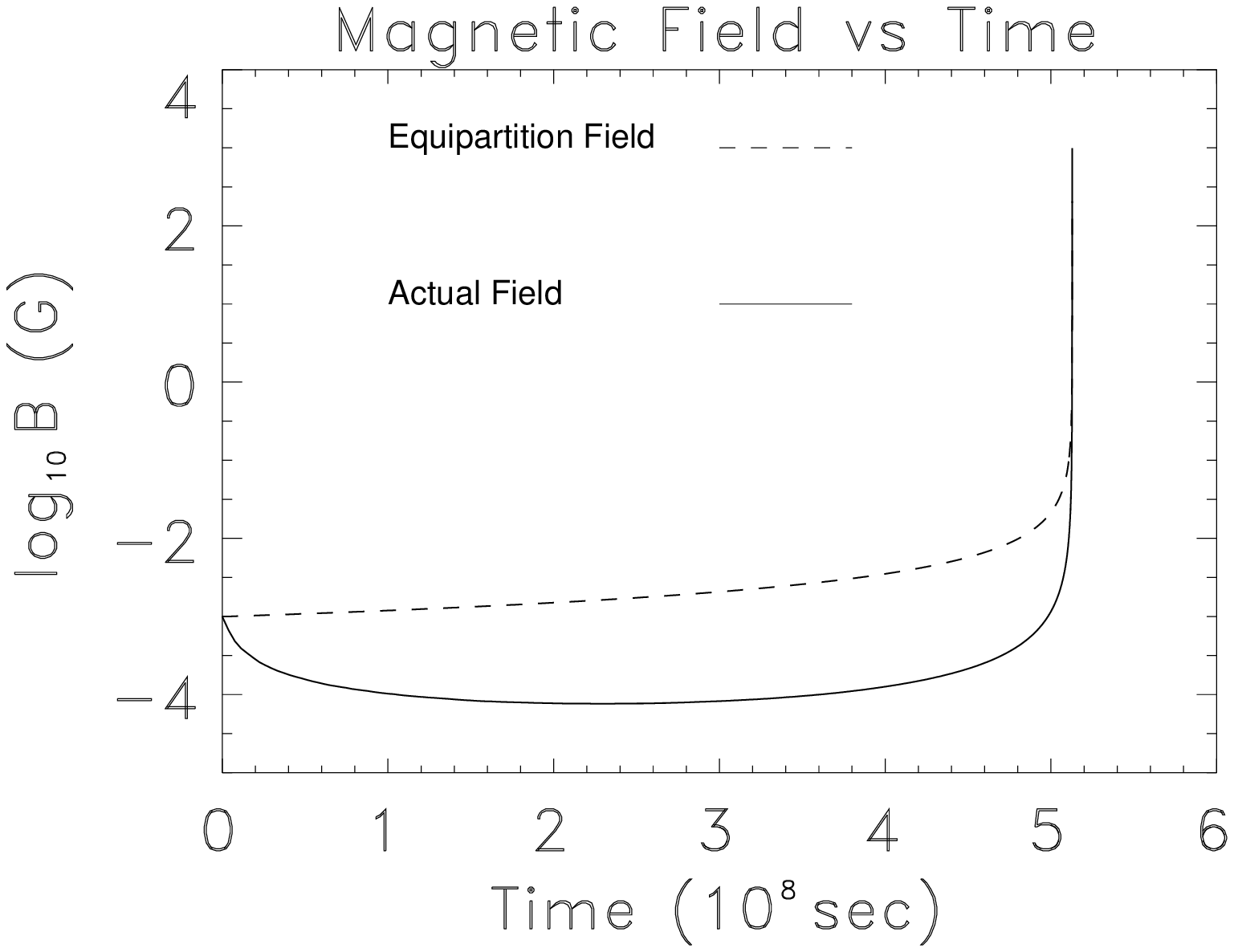,height=12pc}}
\caption{\scriptsize The {\it left} panel shows the magnetic field rates of
change $\partial B/ \partial t|_f$ and $\partial B/\partial t|_d$ due,
respectively, to flux conservation and resistive dissipation as functions of
time in units of $10^8$ seconds.  The compression rate is here
assumed to be the free-fall velocity at the accretion radius.
Solid curve: the rate of increase due to flux conservation;
dashed curve: the rate of decrease due to resistive dissipation.  The {\it right}
panel shows the magnetic field (solid curve) calculated as a function
of time from the rates displayed in left panel.  By comparison,
the equipartition field $B_{eq}$ is shown here as a dashed curve.
The rapid increase in $B$ toward the end of the simulation is associated with
the accelerated rate of change in the physical parameters as the gas flows
inwards toward a zero radius.  (From Kowalenko \& Melia 1999.)\label{magdissipation}}
\end{figure}

Kowalenko \& Melia (1999) have used the van Hoven prescription to calculate
the magnetic field annihilation rate in a cube of ionized gas being compressed
at a rate commensurate with that expected for free-fall velocity onto the
nucleus at the Galactic Center.  An example of these simulations is shown in 
Figure \ref{magdissipation}, for parameter values like those pertaining to the Galactic 
Center.  Whereas the rate of increase $\partial B/\partial t|_f$ in $B$ due to flux 
conservation depends only on the rate $\dot r$ of the gas, the dissipation rate 
$\partial B/\partial t|_d$ (based on the van Hoven prescription) is a function
of the state variables and is therefore not necessarily correlated with
$\dot r$.  Although these attempts at developing a physical model for
magnetic field dissipation in converging flows are still rather simplistic,
it is apparent from the test simulations that the equipartition assumption is not
always a good approximation to the actual state of a magnetohydrodynamic flow,
and very importantly, that the violation of equipartition can vary in degree from
large to small radii, in either direction.  
Coker \& Melia (1999) have calculated the cm to mm spectrum produced by a quasi-spherical 
infall in Sgr~A* using its most recently determined mass, and an empirical fit 
to the magnetic field based on these simulations of magnetic dissipation.  
Without the additional suppression for the radiative efficiency provided by, e.g.,
a two-temperature flow, the implied magnetic field intensity in Sgr~A* is limited 
to a value of about $5-10$ Gauss.  

If Sgr A*'s spectrum is indeed produced by the infalling plasma (as opposed
to an outflowing jet, which we consider in the subsequent section), the geometry of
the emitting region ought to be tightly constrained by the new polarization
measurements described in \S\ 3.7 above.  Although the upper limits to the 
linear polarization in Sgr~A* are found to be quite low (less than $1\%$)
below 86 GHz (Bower et al.~1999), this is not the case at
750, 850, 1,350, and 2,000 $\mu$m, where a surprisingly large intrinsic
polarization of over $10\%$ has now been reported (Aitken, et al.~2000).
These observations also point to the tantalizing result that the position
angle changes considerably (by about $80^\circ$) between the mm and the sub-mm
portions of the spectrum, which one would think must surely have something to do 
with the fact that the emitting gas becomes transparent at sub-mm wavelengths
(Melia 1992, 1994).

\citeN{Agol2000} constructed a simple two-component model for the radio-to-millimeter 
spectrum and the polarization in this source. His analysis predicts that the 
polarization should rise to nearly 100\% at shorter wavelengths. 
The first component, possibly a black hole-powered jet, 
is compact, of low density, and is self-absorbed near 1 mm, with an 
ordered magnetic field, a relativistic Alfv\'en speed, and a 
non-thermal electron distribution. In his model, the second 
component is poorly constrained,
but may be a convection-dominated accretion flow with $10^{-9}\;M_\odot$ yr$^{-1}$, 
in which feedback from accretion onto the black hole suppresses the accretion 
rate at larger radii. This is consistent with the result of
\citeN{QuataertGruzinov2000b}, who show that a high-accretion rate
ADAF would completely depolarize Sgr A*.

\citeN{MeliaLiuCoker2000} have suggested that the mm to sub-mm ``excess" in the 
spectrum of Sgr~A* (see Fig.~\ref{sgrspec}) may be the first indirect evidence 
for the anticipated circularization of the gas falling into the black hole at 
$5-25\;r_s$.  In their simulation of the Bondi-Hoyle
accretion onto Sgr~A* from the surrounding winds, Coker \& Melia (1997) concluded
that the accreted specific angular momentum $l\equiv \lambda r_s c$ can 
vary by $50\%$ over ${\lower.5ex\hbox{$\; \buildrel < \over \sim \;$}}$
200 years with an average equilibrium value in $\lambda$ of about $30$ or less.
The fact that $\lambda\not=0$ therefore raises the expectation that the plasma must 
circularize toward smaller radii before flowing through the event horizon.  
Melia, Liu \& Coker (2000, 2001)
\nocite{MeliaLiuCoker2000,MeliaLiuCoker2001}
showed that this dichotomy, comprising a quasi-spherical
flow at radii beyond $50\; r_s$ or so, and a Keplerian structure toward smaller
radii, may be the explanation for Sgr~A*'s spectrum, including the appearance of
the ``excess", which is viewed as arising primarily within the circularized component. 

\begin{figure}[thb]
\centerline{\begin{turn}{-90}\psfig{figure=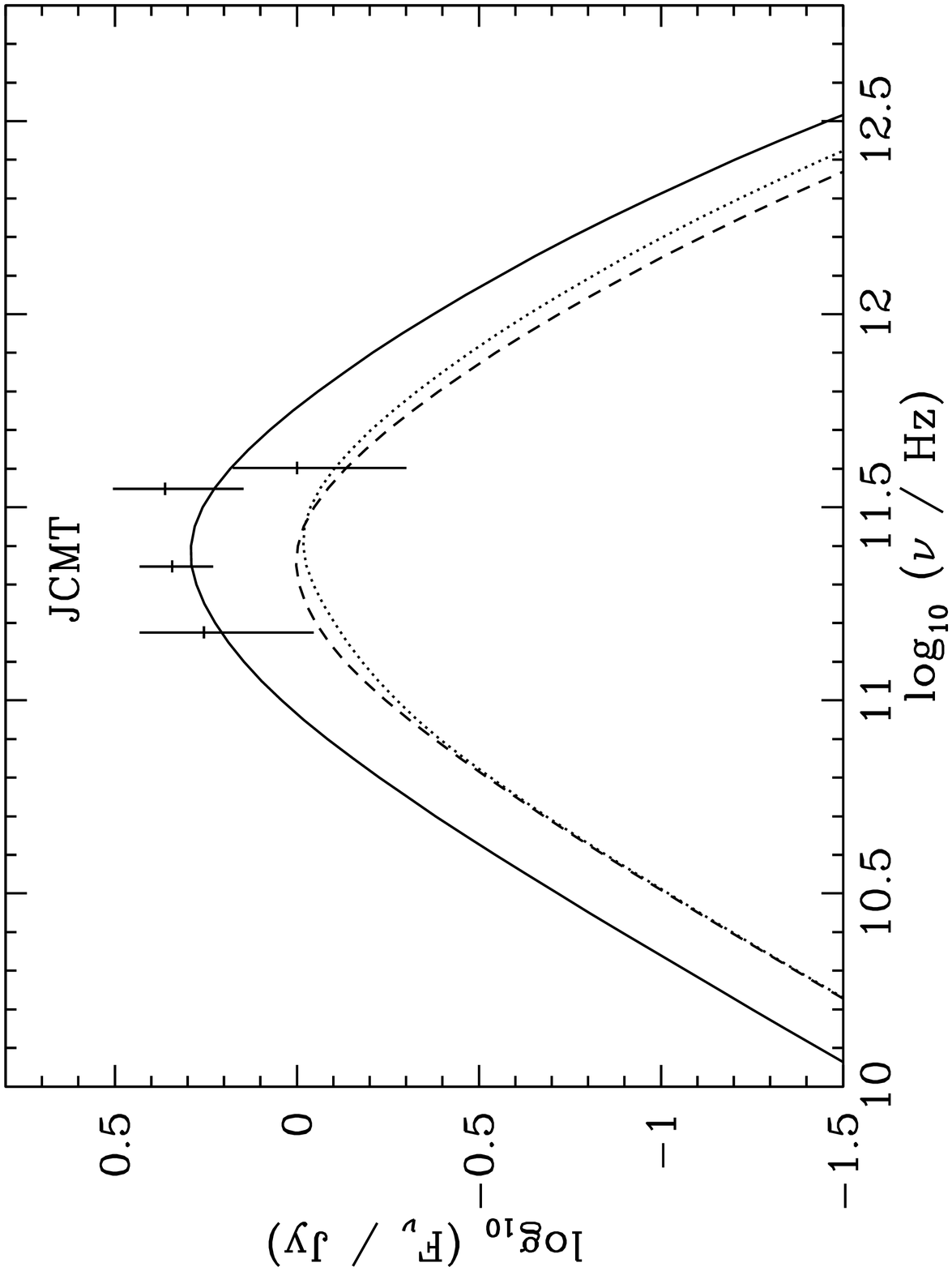,height=16pc}\end{turn}
\begin{turn}{-90}\psfig{figure=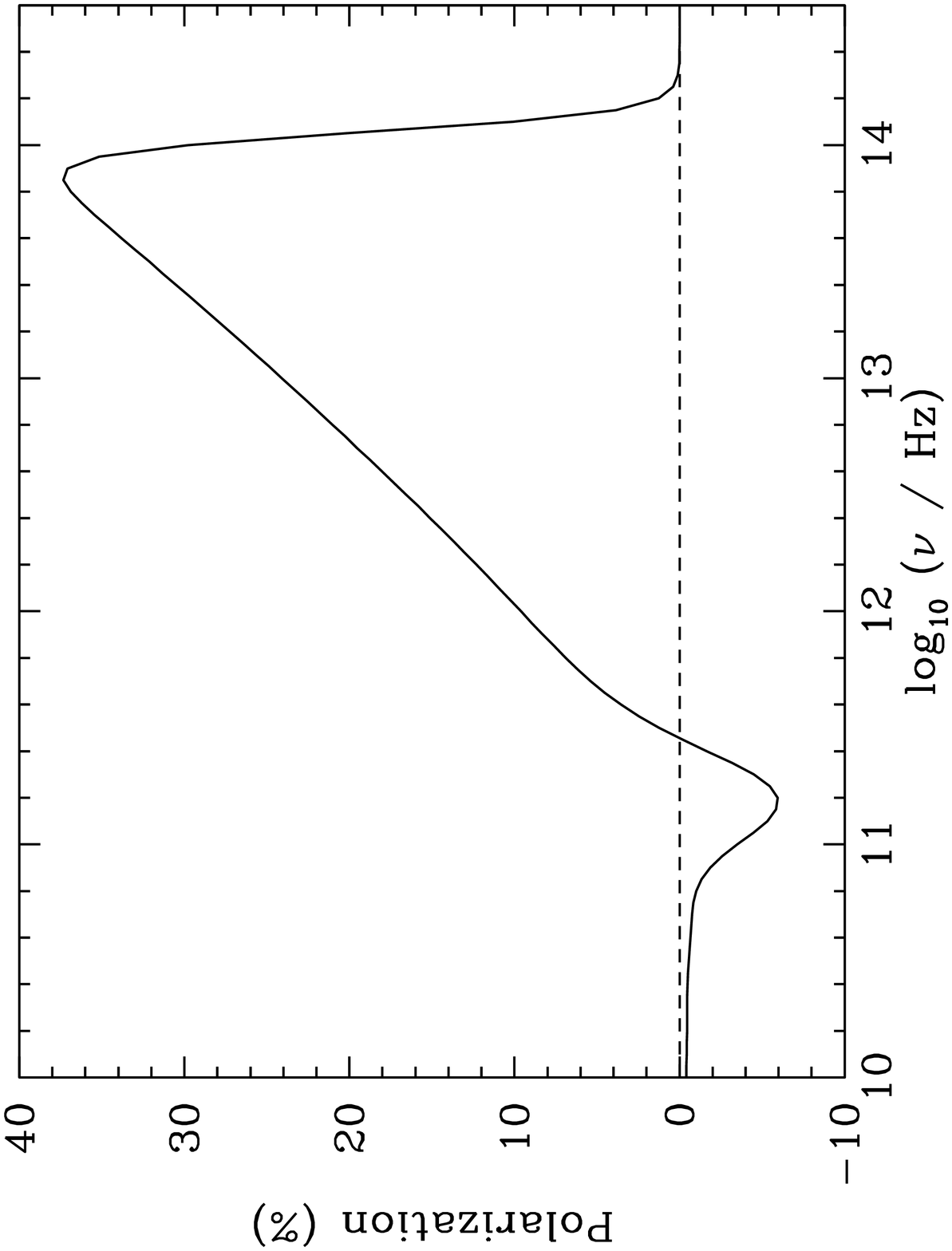,height=16pc}\end{turn}}
\caption{\scriptsize The {\it left} panel shows the spectrum corresponding to the best 
fit model. The dotted curve corresponds to the first component and the dashed
curve corresponds to the second component. The solid curve is the sum of these two. 
The {\it right} panel shows the percentage polarization for this best fit model.
(From Melia, Liu \& Coker 2000.)\label{polarization}}
\end{figure}

In the best-fit model for the polarized mm and sub-mm emission from Sgr~A*, 
the peak frequency of the flux density is $2.4\times 10^{11}$ Hz, and the flip 
frequency (at which the position angle changes by $90^\circ$) is $2.8\times 10^{11}$ Hz. 
Above the peak, the medium is optically thin; it is optically thick at frequencies
below it.  To understand how the polarization characteristics arise in this
context, we note that the circularized flow constitutes a magnetic dynamo that
greatly amplifies the azimuthal component of the magnetic field.
The optically thick emission is dominated by emitting elements
on the near and far sides of the black hole, for which the Extraordinary wave has a
polarization direction parallel to the reference axis. In the left
panel of Figure \ref{polarization}, this optically thick component is indicated by the 
dashed curve.  In contrast, the dominant contribution in the thin region comes from 
the blue shifted emitter to the side of the black hole, where the Extraordinary wave
has a polarization direction mostly perpendicular to this axis. This component
is shown as a dotted curve in the left panel of Figure \ref{polarization}.

Another important result of this analysis is that only modest accretion rates 
appear to be consistent with the polarization characteristics of Sgr~A* at mm 
and sub-mm wavelengths.  The emitting region is compact---evidently no larger 
than a handful of Schwarzschild radii.  Yet hydrodynamical simulations suggest 
a higher rate of capture at larger radii (at $\sim10^4\;r_s$ or so).  If
this modeling is correct, this would seem to suggest that $\dot M$ is
variable, perhaps due to the gradual loss of mass with decreasing radius
that we discussed above.

The low value of $\dot M$ ($<10^{16-17}$ g s$^{-1}$) inferred from the 
polarization studies is significantly smaller than the upper limit
already established for this quantity by the X-ray and IR constraints 
\cite{QuataertNarayanReid1999}.  These authors argue that the combination
of a limit on the X-ray bremsstrahlung emissivity at large radii and the 
IR emissivity from a thick disk at smaller radii, favor an accretion rate
no bigger than about $8\times 10^{-5}\;M_\odot$ yr$^{-1}$, comparable
to the Bondi-Hoyle estimates for the accretion at larger radii
\cite{CokerMelia1997}.

\begin{figure}[thb]
\centerline{\psfig{figure=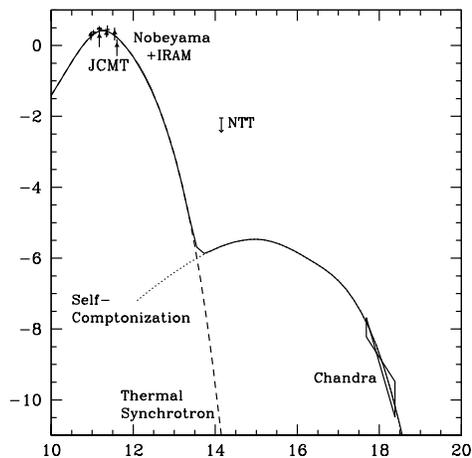,height=16pc}}
\caption{\scriptsize The Comptonized spectrum calculated self-consistently with the
best-fit sub-mm model for the combined Nobeyama and IRAM data gathered in October
1996, and the JCMT data gathered in 1999.  The inferred accretion rate for
this case is $3.5\times 10^{15}$ g s$^{-1}$. The {\it Chandra} data were obtained with 
an observation in 1999.  Dashed curve: thermal synchrotron; Dotted curve: the 
self-Comptonized component; Solid curve: the total spectrum.  The references for 
the radio and X-ray data are given in the text.  Triangles: JCMT data; Bars: Nobeyama 
and IRAM data. (From Melia, Liu, \& Coker 2000.)\label{chandra}}
\end{figure}

The latest {\it Chandra} observations reduce the X-ray limited accretion rate
considerably since for low accretion rates, the dominant contribution to the $0.5-10$ 
keV flux is self-Comptonization within the radio emitting plasma, rather than bremsstrahlung.
The first epoch data show a point source at the location of the central engine 
with a rather low X-ray luminosity ($\sim [0.5-0.9]\times10^{34}$ erg s$^{-1}$) 
in the 0.5-10 keV band \cite{BaganoffBautzBrandt2001}.
Figure \ref{chandra} shows the complete spectrum that includes the thermal
synchrotron emission in the sub-mm range, together with the self-Comptonized
component \cite{MeliaLiuCoker2000}. These authors find that a best fit
to the Nobeyama and IRAM data alone produces a corresponding X-ray flux that is
too high (compared to the {\it Chandra} measurement) by about a factor of $4$,
whereas a best fit to the JCMT data produces a self-Comptonized flux that is
too low by the same factor.  The fit shown in this figure is for the combined
sub-mm data sets. (The NTT upper limit is from \citeNP{MentenReidEckart1997}.)  
This may be interpreted as an indication of the source 
variability (in both the sub-mm and X-ray portions of the spectrum) between 
1996 and 1999. More specifically, the accretion rate through the inner Keplerian
region appears to have decreased by about $15\%$ between the two radio measurements.
The implied correlated variability between the sub-mm and X-ray fluxes 
suggests that future observations with {\it Chandra} may directly test this basic
picture for the sub-mm to X-ray emissivity in Sgr A*. 

It is clearly essential to now self-consistently match the conditions within 
the Keplerian region of the flow with the quasi-spherical infall further out.  
These calculations are necessary and timely.  In particular, it is
important to update the early estimates for the frequency-dependent size
of Sgr A* \cite{MeliaJokipiiNarayanan1992} in view of the improved
restrictions on the properties of the accreting gas.  At each frequency,
the emission is dominated by the ``shell" of gas at which the radiation 
becomes self-absorbed, thus stratifying the medium.  This can provide a 
powerful tool for testing our understanding of this system with 
frequency-dependent imaging of the central region. 

In summary, the emission in Sgr~A* (if produced in the accretion region)
requires a very deep potential well, so 
the case for a massive black hole rather than a distributed dark matter has 
grown stronger.  Whether the radiation mechanism is thermal or non-thermal, 
the radiative efficiency of the infalling gas must be very low ($< 10^{-5}$).
All things considered, this low efficiency is probably due to either a 
sub-equipartition magnetic field (for either thermal or non-thermal models), 
or to the separation of the gas into a two-temperature plasma with $T_e \ll T_p$.  
The current limit on the accreted specific angular momentum appears to be
inconsistent with the formation of a large disk, fossil or otherwise
\cite{FalckeMelia1997}, favoring instead the circularization of the infalling 
plasma when it plummets to within $50\,r_s$ or so of the black hole. The spectral 
and polarization data pertaining to the sub-mm bump are consistent with this 
portion of the spectrum arising from the inner Keplerian flow within $10-20\,r_s$ of
the accretor. 

\subsection{Emission due to Non-Accretion Processes}\label{jetmodel}
If the dominant emission is not due to the accreting gas one can
consider whether the observed spectrum is in fact produced by an
outflow or a jet launched from the vicinity of the black hole. The 
essential element of this model is the ejection of a magnetized plasma 
containing relativistic electrons or pairs through a nozzle above the 
event horizon. In this picture, the infalling plasma emits rather
weakly but may contribute to the expulsion of matter which is then 
responsible for the radio and X-ray emission.

In the context of AGNs, \citeN{BlandfordKonigl1979} proposed that 
flat-spectrum radio cores are incoherent, nonthermal, synchrotron-emitting
jets. Stimulated by this, \citeN{ReynoldsMcKee1980} first considered
the possibility that the flat to inverted radio spectrum of Sgr~A*
may be due to an analogous jet or wind from a stellar object or a supermassive
black hole. They also argued that Sgr~A* is unlikely to be
gravitationally bound, because its equipartition energy density is too
large. The idea was revived by \citeN{FalckeMannheimBiermann1993} who,
based on the jet-disk symbiosis idea \cite{FalckeBiermann1995,FalckeBiermann1999},
showed that the basic properties of Sgr A* may be explained by a
scaled down AGN-jet model, requiring a very low accretion rate 
($\dot M \ga 10^{(-8)-(-7)} M_\odot$ yr$^{-1}$) to power the outflow.

Possible launching mechanisms for the jets have been discussed in the literature
(e.g., \citeNP{ApplCamenzind1993,KoideMeierShibata2000}) and usually
invoke pressure and magnetohydrodynamic acceleration of plasma from
the inner edge of an accretion flow. This acceleration region close to
the black hole, summarily called a ``nozzle", should radiate at the
highest synchrotron frequencies, i.e., in the sub-mm regime for Sgr A*.
However, as realized by \citeN{DuschlLesch1994}, the steep cut-off
towards the IR in the Sgr A* spectrum requires a rather narrow energy
distribution for electrons, i.e., a quasi-monoenergetic population.
Such a distribution is consistent with a Maxwellian, which was
shown earlier to result in a transparent medium (i.e., a sharp
drop-off in flux) above $\sim (2-3)\times 10^{11}$ Hz \cite{Melia1992a,Melia1994}.
Self-absorption, on the other hand will occur somewhere in
the mm-wave regime and thus, together with the peaked electron
distribution, produces a peaked spectrum \cite{Falcke1996b,BeckertDuschl1997}.

To capture the basic elements of this picture, we may consider
a simple toy model with four parameters: the magnetic field $B$,
the electron density $n$, the electron Lorentz factor $\gamma_{\rm e}$, and
the volume $V=4/3\,\pi R^3$, using for simplicity a one-temperature
(i.e., a quasi mono-energetic) electron distribution, with the distance
being set to $8.5$ kpc.  On the observational side we have three measurable
input parameters: the peak frequency $\nu_{\rm max}\sim\nu_{\rm
c}/3.5$ of the synchrotron spectrum (characteristic frequency
$\nu_{\rm c}$), the peak flux $S_{\nu_{\rm max}}$, and the
low-frequency turnover of the sub-mm bump at the self-absorption
frequency $\nu_{\rm ssa}$. A fourth parameter can be gained if one
assumes that the magnetic field and relativistic electrons are in approximate
equipartition, i.e.~$B^2/8\pi=k^{-1} n_{\rm e} \gamma_{\rm e} m_{\rm
e} c^2$ with $k\sim1$. With this condition one obtains from
synchrotron theory that

\begin{equation}\label{submmpars1}
\gamma_{\rm e}=118 \; k^{2/7}
\left({\nu_{\rm max}\over {\rm THz}}\right)^{5/17}
\left({\nu_{\rm ssa}\over 100 {\rm GHz}}\right)^{-5/17}
\left({F_{\nu_{\max}}\over 3.5 {\rm Jy}}\right)^{1/17},
\end{equation}

\begin{equation}
B=75\,{\rm G}\; k^{-4/17}
\left({\nu_{\rm max}\over {\rm THz}}\right)^{7/17}
\left({\nu_{\rm ssa}\over 100 {\rm GHz}}\right)^{10/17}
\left({F_{\nu_{\max}}\over 3.5 {\rm Jy}}\right)^{-2/17},
\end{equation}

\begin{equation}
n_{\rm e}=2\times10^6 {\rm cm^{3}}\;k^{7/17}
\left({\nu_{\rm max}\over {\rm THz}}\right)^{9/17}
\left({\nu_{\rm ssa}\over 100 {\rm GHz}}\right)^{25/17}
\left({F_{\nu_{\max}}\over 3.5 {\rm Jy}}\right)^{5/17},
\end{equation}

\begin{equation}\label{submmpars2}
R=1.5\times10^{12} {\rm cm}\;k^{-1/17}
\left({\nu_{\rm max}\over {\rm THz}}\right)^{-16/51}
\left({\nu_{\rm ssa}\over 100 {\rm GHz}}\right)^{-35/51}
\left({F_{\nu_{\max}}\over 3.5 {\rm Jy}}\right)^{8/17}
\end{equation}

Apparently the parameter $k$ enters only weakly
and hence the above values should reflect the characteristic
properties of the sub-mm emission region in Sgr A* to within a factor
of a few. The electron Lorentz factor corresponds to around
$2\times10^{11}$ K using $\gamma_{\rm e}\sim\sqrt{12}k_{\rm b} T_{\rm
e} (m_{\rm e} c^2)^{-1}$ the average Lorentz factor for a relativistic 
Maxwellian (see \S\ 5.1). The size $R$ corresponds to $\sim2$ Schwarzschild
radii and is consistent with the sub-mm bump coming from the direct
vicinity of the black hole.

If these parameters describe the nozzle of a jet, rather than a static
corona as proposed by \cite{BeckertDuschl1997},
the emission at lower frequencies can be obtained in a straight
forward manner by following the evolution of the plasma on its way out
using the Euler equation. In the
supersonic, post-nozzle regime the jet mainly accelerates through its
longitudinal pressure gradient to bulk Lorentz factors around
$\gamma_{\rm j}=2-3$. As the plasma moves outwards and expands,
roughly filling a conical jet with $B\sim r^{-1}$ and $n\sim r^{-2}$,
the peak synchrotron frequency will drop continuously. Integration
over the entire length of the jet, taking into account the changing
Doppler factor and adiabatic losses yields a slightly inverted radio spectrum
\cite{Falcke1996a} with $\alpha\simeq0-0.25$ as a function of the
inclination angle.

\begin{figure}
\centerline{\psfig{figure=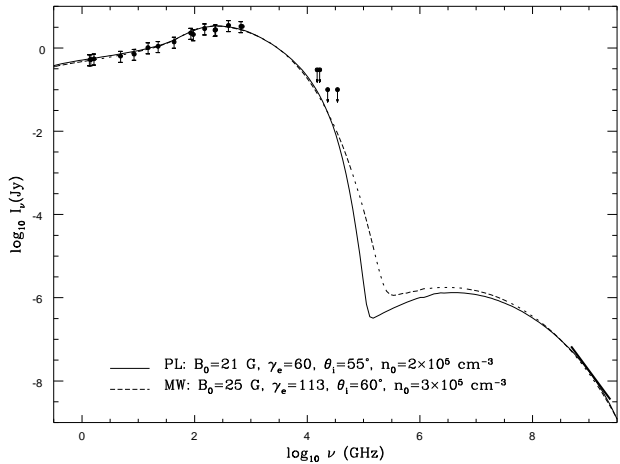,width=0.6\textwidth,angle=-90}}
\caption[]{\label{sgrx-pl}\scriptsize
Broad-band spectrum of Sgr~A* produced by a jet model, with a power-law 
electron distribution (PL) and a relativistic Maxwellian (MW). The
width of the nozzle is $r_0=4\,r_s$ and $r_0=3 r_s$, respectively, 
while its height is at $z_0=3r_0$. The data are from
Fig.~\ref{sgrspec}, augmented by the high-frequency measurements 
discussed in \citeN{SerabynCarlstromLay1997}. The {\it Chandra} 
data are also shown. (From \citeNP{FalckeMarkoff2000}.)}
\end{figure}

\citeN{FalckeMarkoff2000} have carried out these calculations for
arbitrary electron energy distributions. Figure \ref{sgrx-pl} shows
the results for a Maxwellian and a curtailed power-law
distribution. The model can account for the cm to sub-mm
spectrum. Moreover, when one calculates synchrotron radiation from
the relativistic electrons, one also has to take into account that the
very same electrons will up-scatter their own synchrotron photons
via the synchrotron self-Compton process (SSC), as was the case for the infall
model (see \S\ 5.1). Since the SSC emission is proportional to
$n_{\rm e}^2$ the emission is dominated by the most compact region,
in this case, the sub-mm bump or the nozzle. For electron
Lorentz factors $\gamma_{\rm e}\sim10^2$ the scattered sub-mm emission
should then appear at very soft X-rays. Indeed, for Sgr~A* the SSC
component appears as a second bump in Figure
\ref{sgrx-pl}, accounting for the low and very soft X-ray emission
detected by Chandra \cite{BaganoffBautzBrandt2001}. If this
interpretation is correct one would expect to see correlated
variability between sub-mm and X-ray emission, though it is not
yet clear how the variability amplitudes compare with those
of the accretion model (see Fig.~\ref{chandra}).

As is the case in any stratified emission model, the radiation from the jet at
different frequencies will be dominated by different regions where the
optical depth $\tau$ approaches unity. This yields roughly a $\nu^{-0.9}$
dependence of the characteristic size of the emission region as seen
for example in extragalactic sources like M81* \cite{BietenholzBartelRupen2000}.  
This also predicts a frequency-dependent core-shift which
should be observable with VLBI phase-referencing observations. The
peaked electron distribution and the absence of a power-law at high
frequencies in Sgr A* also imply that the emission at one observing
frequency is very compact highlighting only a narrow section of the
jet. This is consistent with current VLBI observations.

The narrow electron distribution needed in Sgr A* is rather unusual
and may indicate the absence of shocks along the jet commonly found in
more luminous AGNs. If the jet is launched from a two-temperature plasma, 
the electrons would not be hot enough while the protons could reach temperatures in
excess of $10^{12}$ K at the inner edge as a function of black hole
spin \cite{Manmoto2000}. At this temperature proton-proton collisions
will become inelastic and lead to the production of pions
with a subsequent decay into neutrinos, $\gamma$-rays, and enough
energetic pairs with $\gamma_{\rm e}\simeq60$ to account for the jet
emission \cite{MarkoffFalckeBiermann2001}. The jet plasma in such a
model would then be a mixture of ``cold" electrons and ``hot" pairs. If the
number of cold electrons from the normal plasma dominates, this in
itself could lead to substantial depolarization of linear polarization
and conversion to circular polarization \cite{BowerFalckeBacker1999,BeckertFalcke2001}.

In future work, it will be necessary to understand how the jet, 
if present, is coupled to the inflowing plasma (see, e.g., 
\citeNP{Yuan2000}).  Is the accreting gas responsible
for producing it, or is the jet simply a by-product of a spinning
black hole?  And in either case, if we're seeing emission from
a jet in Sgr A*, why is the accreting plasma so underluminous? 

This latter condition is also a necessary feature in models that
invoke static configurations of hot gas.  \citeN{DuschlLesch1994}
and \citeN{BeckertDuschl1997} explain the radio to FIR-spectrum of 
Sgr A* as incoherent, optically thin synchrotron radiation from 
relativistic electrons (and/or positrons) bound to the central
gravitational potential.  If the spectrum is treated in a time-averaged
fashion (so that it has a dependence $\sim \nu^{1/3}$ between $1$ 
and $10^3$ GHz) the required particle distribution is quasi-mono energetic with 
$\Delta E/E < 7$ ($E$ being the characteristic electron energy and $\Delta 
E$ the width of its distribution). \citeN{BeckertDuschl1999} consider relativistic 
thermal distributions as a natural subclass and an acceptable fit to the 
time averaged spectrum. Acceleration processes that may lead to such 
quasi-monoenergetic distributions of electron energies are discussed by 
\citeN{DuschlJauch2000}.

In this picture, Sgr A* is modeled as a core-shell structure with two homogeneous 
components. The core with an electron temperature of $T_e = 5\times 10^{11}$ K and a 
magnetic field of $B = 70$ G is only marginally larger than the Schwarzschild radius, 
and is visible only as the sub-mm excess flux. In a variation of this basic
model, \citeN{BeckertDuschl1999} discuss a core component made up of the central 
regions of an ADAF disk. The much more extended shell is optically thin above 
$\approx 2$ GHz and is filled with electrons of $T_e = 2\times 10^{12}$ K in a 
$2$ G magnetic field. The self-Comptonization of synchrotron photons by the 
relativistic electrons is minimal due to the small Thomson optical depth of 
$\tau \approx 10^{-2}$ and appears in the UV and soft X-rays. The corresponding
flux can be matched to the {\it Chandra} measurement, but the spectral shape 
in X-rays depends strongly on the column density of absorbing material and the 
electron temperature in the core component.

Future work with this model will need to address issues such as (1) how does
the static configuration account for the {\it instantaneous} spectrum of Sgr A*?
(2) what determines the temperature and magnetic field of the plasma?
(3) what are the implied polarization properties of this gas? and (4)
what produces the variability at radio frequencies?  It now appears that some of 
the short-term variability in Sgr A* is indeed intrinsic to the source and a 
variable spectral index appears to be incompatible with a one- or two-component model.

In conclusion, we can say that, while the underlying concepts for the
various emission models of Sgr A* sound rather different, physical 
quantities such as $B$, the particle density, and the temperature and/or particle
Lorentz factor, have values that are slowly converging with one another.
This is due primarily to the ever improving observational constraints and it
is expected that the current degeneracy of models may collapse to a
unified picture involving some of the ideas discussed in these sections.

\subsection{Alternatives to the Black Hole Paradigm}
Alternatives to a supermassive black hole as the central dark mass
concentration have so far tended to concentrate on the structure of
the central object, rather than its emissivity.  Nonetheless, the 
current observational limits inferred from Sgr A*'s spectrum do 
tightly constrain (or even exclude) some of them.  However, until definitive
proof of an event horizon has been obtained (see \S\ \ref{GR}),
other possibilities must remain open---indeed should continue to be
explored.

One idea that has been explored recently is that of a nonbaryonic ball
comprised of degenerate, self-gravitating heavy neutrino matter
\cite{TsiklauriViollier1998}. Its size is a strong function of the 
neutrino mass $m_\nu$.  This scenario requires us to postulate the
existence of as yet unidentified neutrinos with mass $> 17$ keV,
which would condense into a sphere with a characteristic size of
about $0.01$ pc. However, to explain even more massive dark compact objects, 
such as that in M87 with $10^9\,M_\odot$, the putative neutrino mass cannot 
be greater than 17 keV, for otherwise the neutrino ball itself attains
an event horizon.  Given that the new stellar orbits determined by 
\citeN{GhezMorrisBecklin2000a} at the Galactic Center limit the volume of 
the region within which the dark matter is contained even further, 
the neutrino mass would now need to be substantially greater than this
limit. 

It thus appears that a black hole-free universe probably cannot
be constructed in this way, but let us suppose that Sgr A* is a 
neutrino ball. Its luminosity in that case would be due to disk
emission from gas spiraling through the gravitational potential of
a radially-dependent enclosed mass. Thus, at any given radius, 
the dissipation rate falls below the corresponding value for the case
where all the mass is concentrated at a central point.  This introduces
the attractive feature of accounting for a decreasing radiative
efficiency as the gas approaches the middle.  However, it also begs
the question of what happens to the infalling matter.  Presumably,
it stays trapped within the neutrino ball, but over the age of the 
Galaxy, some $10^6\,M_\odot$ of plasma will have condensed to the 
bottom of the potential well, assuming that Bondi-Hoyle accretion 
proceeds at the rate suggested by the large scale simulations
(see above).  In a sense, this defeats the purpose of having
a ball of degenerate neutrinos.

Of course, one could consider other particles to reconstruct the
dark compact mass at the Galactic Center.  Instead of fermions,
one could try bosons, such as Higgs particles, and postulate 
a massive boson star, perhaps due to topological defects in the 
cosmological evolution \cite{TorresCapozzielloLambiase2000}. However,
at this writing, these models remain very speculative and no clear bound 
on the required particle mass can be given. On the other hand, supermassive 
stars of ordinary matter with an even heavier accretion disk \cite{Kundt1990} 
can already be comfortably ruled out, based on the low near-infrared flux at
the position of Sgr A*. Finally, suggestions that Sgr A* could be a
matter-creating ``white hole'' \cite{BurbidgeHoyle1996}
or an accreting ``near-black-hole" need to be fleshed out with more detail, 
commensurate with the richness of the current database for this object.

\section{STRONG GRAVITY EFFECTS}\label{GR}
The ever growing interest in Sgr~A* has already yielded a number of
tantalizing results, the most important being that Sgr~A* is the best
supermassive black hole candidate we know.  Beyond trying to model
the emission from this source, it is worth thinking about the possibility
of utilizing its relative proximity in order to test the predictions of 
General Relativity in the strong field limit.  For example, the fact
that Sgr A*'s mass is known so precisely and that the emitting gas
is apparently becoming transparent at mm to sub-mm wavelengths near
the marginally-stable orbit, means that timing studies of this
source with bolometric detectors on single-dish telescopes
may reveal the black hole's spin \cite{MeliaBromleyLiu2001}.
Surprising as this may
seem, we are at the stage where we can begin to ask questions such
as ``Is there really an event horizon in this source?''  Embedded within
a bright star cluster, Sgr A* might also be a microlens, producing
effects that will be measurable with our ever improving spatial
resolution of this region. 

\subsection{Imaging the Event Horizon}\label{eventhorizon}
The VLBI resolution is rapidly approaching a scale commensurate with the 
actual size of Sgr A*'s event horizon.  When we realize that the presence 
of the sub-mm bump in the spectrum is indicative of a compact 
emission region a mere couple of Schwarzschild radii in size,
it becomes worthwhile exploring the possibility of actually ``seeing" 
the shadow of the black hole using VLBI imaging techniques. This 
naturally will have to be done at the highest radio frequencies where 
the resolution is the best, and the scatter-broadening of Sgr~A* 
by the intervening medium is the lowest.

At sub-mm wavelengths, the synchrotron emission is not self-absorbed
\cite{Melia1992a,Melia1994,Falcke1996b}; the medium's transparency at the shortest 
wavelengths allows us to view the emitting gas all the way down to the 
event horizon, whose size is $(1+\sqrt{1-a_*^2})r_s/2$, where $r_s\equiv 
2GM/c^2$, $M$ is the mass of the black hole, $G$ is Newton's constant, 
$c$ the speed of light, $a_*\equiv Jc/(GM^2)$ is the dimensionless spin 
of the black hole in the range 0 to 1, and $J$ is the angular momentum of 
the black hole.  \citeN{Bardeen1973} described the idealized appearance 
of a Schwarzschild black hole {\it in front} of a planar emitting source 
(e.g., a star), showing that it literally would appear as a ``black hole" 
of diameter $\sqrt{27} r_s/2$. At that time, such a calculation was of mere 
theoretical (rather than practical) interest. To further check
whether there is indeed a realistic chance of seeing this ``black hole"
in the Galactic Center, \citeN{FalckeMeliaAgol2000} simulated the
appearance of the emitting gas {\it surrounding} Sgr~A* using a
general relativistic (GR) ray-tracing code for various combinations of
black hole spin, inclination angle, and morphology of the surrounding
emission region. The simulations take the scatter broadening and the
instrumental resolution of VLBI at sub-mm waves into account.

As revealed by these calculations the presence of an event horizon
inside a transparent radiating source will naturally lead to a deficit
of photons in the center, called a ``shadow" by \citeN{FalckeMeliaAgol2000} 
and independently \citeN{deVries2000}. The size of the shadow is larger 
than the event horizon due to the strong bending of light by the black 
hole and is of order $5r_s$ in diameter.

\begin{figure}[thb]
\centerline{\psfig{figure=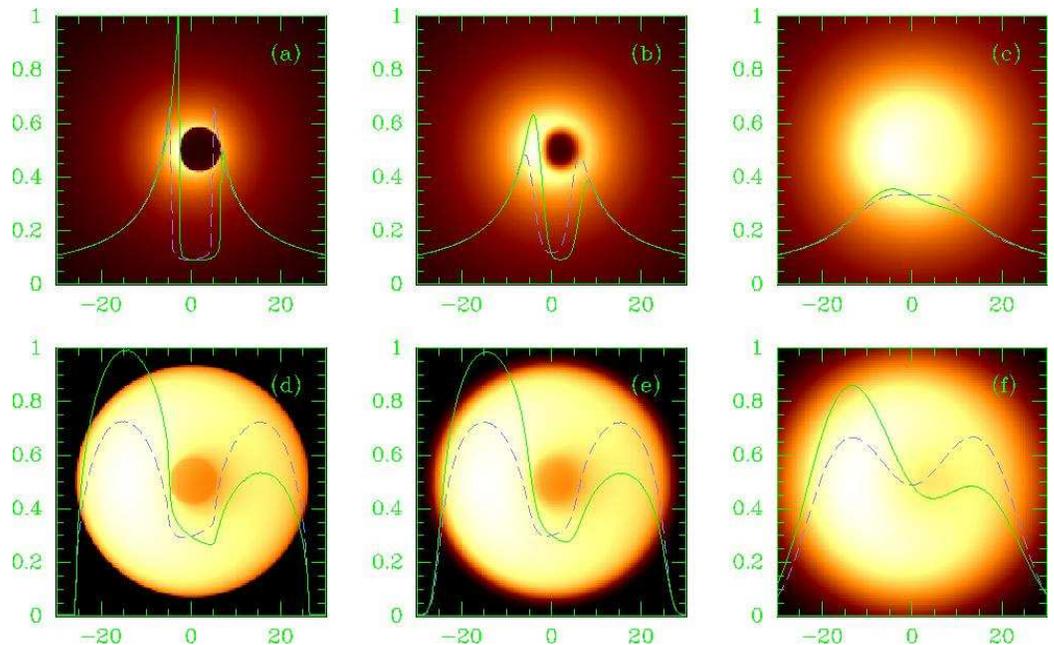,height=20pc}}
\caption{\scriptsize An image of an optically thin emission region 
surrounding a black hole with the characteristics of Sgr~A* at the 
Galactic Center.  The black hole is here either maximally rotating 
($a_* = 0.998 $, panels a-c) or non-rotating ($a_*=0$, panels d-f). 
The emitting gas is assumed to be
in free fall with an emissivity $\propto r^{-2}$ (panels a-c) or on Keplerian
shells (panels d-f) with a uniform emissivity (viewing angle
$i=45^\circ$). Panels a\&d show the GR ray-tracing calculations,
panels b\&e are the images seen by an idealized VLBI array at 0.6 mm
wavelength taking interstellar scattering into account, and panels
c\&f are those for a wavelength of 1.3 mm. The intensity variations
along the $x$-axis (solid green curve) and the $y$-axis (dashed
purple/blue curve) are overlayed. The vertical axes show the intensity
of the curves in arbitrary units and the horizontal axes show the
distance from the black hole in units of $r_s/2$ which for Sgr~A*
is $3.9\times 10^{11}$ cm $\sim3\;\mu$as. 
(From Falcke, Melia \& Agol 2000.)\label{bhimage}}
\end{figure}

Two disparate cases are reproduced here (see Fig.~\ref{bhimage}),
which include a rotating and a non-rotating black hole, a rotating and
an inflowing emission region, as well as a centrally peaked and a uniform 
emissivity.

The shadow can be clearly seen with a diameter of $4.6\,r_s$ (30
$\mu$as) in diameter for the rotating black hole and with a radius of
$5.2\,r_s$ (33 $\mu$as) for the non-rotating case. The emission
can be asymmetric due to Doppler shifts associated with rapid rotation
(or inflow/outflow) near the black hole. The size of this shadow is
within less than a factor of two of the maximum resolution already 
achieved by sub-mm VLBI ($\sim50\mu$as, \citeNP{RantakyroBaathBacker1998}).
It may also be feasible to do polarimetric imaging at mm and sub-mm
wavelengths, which would reveal additional effects of strong gravity
distortions \cite{BromleyMeliaLiu2001}.

Interestingly, the scattering size of Sgr~A* and the resolution of
global VLBI arrays become comparable to the size of the shadow at a
wavelength of about 1.3 mm. As one can see from Figure \ref{bhimage},
the shadow is still almost completely washed out for VLBI observations
at 1.3 mm, while it is very apparent at a factor two shorter
wavelength (Figs.~\ref{bhimage} b\&e). In fact, already at 0.8 mm
(not shown here) the shadow can be seen easily. Under certain
conditions, i.e., a very homogeneous emitting region, the shadow would
be visible even at 1.3 mm.  The technical methods to achieve such a
resolution at wavelengths shorter than 1.3 mm are currently being
developed and a first detection of Sgr~A* at 1.4 mm with VLBI has
already been reported \cite{KrichbaumGrahamWitzel1998}. Pushing the
VLBI technology even further, towards $\lambda$0.8 or even
$\lambda$0.6 mm, should eventually provide the first direct evidence
for the existence of an event horizon. Alternatively one could think
of space-based X-ray imaging of this shadow as has been proposed
recently \cite{CashShipleyOsterman2000}. However, this technology is
still far in the future and Sgr~A* is rather weak in X-rays.

The imaging of this shadow would confirm the widely held belief that 
most of the dark mass concentration in the nuclei of galaxies such as 
ours is contained within a single object. A non-detection with sufficiently 
developed techniques, on the other hand, might pose a major problem for 
the standard black hole paradigm. Because of this fundamental importance, 
this experiment should be a major motivation for intensifying the current 
development of sub-mm astronomy in general and mm- and sub-mm VLBI in particular.

\subsection{Interactions of Sgr A* with the Central Star Cluster}\label{starcluster}
Besides gravitational light bending very close the event horizon, one
might also expect to see microlensing of stars by Sgr A*. When
a star passes behind Sgr A*, we would expect to see a temporary
amplification of its luminosity. These events occur with a rate
that depends strongly on the assumed stellar distribution in the Galactic
Center and in the Galactic plane. Current estimates predict about 
$10^{-3}$ events per year for amplifications lasting one year down to a
detection limit of 17 mag in K \cite{AlexanderSternberg1999}. 
For current telescopes and monitoring programs, the detectability 
of this effect is rather low. However, the probability of 
{\it microlensing} the background stars is in fact increased by 
the combined action of a central black hole and the dense central 
star cluster. Depending on the stellar background density, this effect 
could provide a 1\% probability of seeing a microlensing 
effect in the inner 2\arcsec at the Galactic Center at any given
time \cite{AlexanderLoeb2001}. This is because the Einstein radius of 
Sgr A* for a source at infinity is rather large, i.e., about 1\farcs75.
Since lensed images of a star on the opposite side of a black hole
should be lined up with the black hole itself, one can try to use current
NIR maps to look for such a correlation in a statistical way.  Using
this method, Alexander (2001) finds further evidence
for Sgr A* being coincident with the center of gravity in the Galactic nucleus.

Another impact Sgr A* may have on the surrounding stars is the tidal
disruption of cluster members when they venture too close to the 
black hole \cite{Rees1982,KhokhlovMelia1996}. Again, the
actual event rate depends on the exact stellar density, but is expected
to be around several times $10^{-5}$ yr$^{-1}$ \cite{Alexander1999}. While
it is unlikely for us to directly witness such an event, its remnant
could still be visible today. In this regard, the inferred age of Sgr A 
East, its morphology, and energetics would fit such a scenario 
\cite{KhokhlovMelia1996}. In addition the yet unidentified fossil 
remains of previous explosions might also be visible in low-frequency 
observations of the Galactic Center \cite{KassimLarosaLazio1999}
and in the distribution of electron-positron annihilation radiation
from the central bulge \cite{FatuzzoMeliaRafelski2001}.

While some of these arguments offer supportive, rather than direct,
evidence for the existence of the black hole, future determinations 
of the stellar orbits at the Galactic Center using space interferometry 
(e.g., with {\it DARWIN}; \citeNP{Wilson2000}) promise much more insight. 
These orbits may not help us to distinguish between rotating and non-rotating black
holes, but they will clearly differentiate between point and extended mass 
distributions \cite{MunyanezaTsiklauriViollier1999}.

\section{SGR A* AS A MODEL FOR AGN ACTIVITY}
The prospects for applying what we are learning in the Galactic Center to 
the broader study of Active Galactic Nuclei (AGNs) are quite promising.  
We mention only a few areas of overlap here, but the cross-fertilization 
is likely to blossom quickly into the future.  The recent work on the gas 
dynamics in the Galactic Center has greatly improved our understanding of 
the gaseous flows surrounding a massive black hole, particularly with regard 
to properties such as the specific angular momentum distribution, density, 
and temperature of the inflowing plasma.  With the appropriate extrapolation 
of the physical conditions, this information can be valuable in trying to 
determine the origin of the Broad Line Region (BLR) in AGNs. 
On larger scales, we see the importance and action of magnetic fields. 
While our view is still patchy, we have observed toroidal magnetic fields 
in molecular clouds \cite{Novak1999} accreting towards the center 
\cite{vonLindenBiermannDuschl1993} and interacting with the large
scale poloidal field seen in the filaments 
\cite{Morris1994,ChandranCowleyMorris2000}.

The molecular clouds, especially the circumnuclear disk (CND),
could be the Galactic Center's version of the obscuring torus inferred for
many AGNs \cite{Antonucci1993}. The CND/torus might simply be
the remnant of tidally disrupted clouds \cite{Sanders1999} trapped in
the transition region where the black hole mass starts to dominate the
gravitational potential \cite{Duschl1989}.

Further in we see what could be considered the Narrow-Line-Region of
AGNs: the gas streamers and colliding stellar winds. The hot gas in the minispiral
is a strong emitter of narrow H$\alpha$ (see the NICMOS Pa $\alpha$
image in Fig.~\ref{fig-paa}) and similarly narrow (in AGN terms)
emission lines are produced by the stellar winds from luminous
stars in the center. Currently, the excitation of this gas is only due
to stars and would at best resemble an \ion{H}{2} galaxy
\cite{ShieldsFerland1994}, as found in our cosmic neighborhood
\cite{HoFilippenkoSargent1997a}. However, with a velocity dispersion
of several hundred km s$^{-1}$ this gas would immediately turn into a
typical NLR should Sgr A* light up in the future. The
presence of so much gas near the black hole suggests that 
sooner or later the accretion onto Sgr A* might become much
higher. We may have already undergone the first stage of this 
accretion event in the form of a star burst several million years
ago, producing today's young and hot stars in the central
parsecs.  \citeN{MorrisGhezBecklin1999} have argued on this basis that the
Galactic Center may be exhibiting a limit cycle of recurrent
nuclear activity, with a timescale ($\sim 10^7$ yrs) dictated by
the evolution of the most massive stars. This highlights the 
starburst-AGN connection.

So far we have not seen any evidence for a broad line region near Sgr A*.
The spectra of many AGNs, including Seyfert galaxies and quasars, are 
distinguished by strong, broad emission lines, with a full width at half maximum 
intensity (FWHM) of $\sim 5,000\ \mathrm{km\ s^{-1}}$, and a full width at 
zero intensity (FWZI) of $\sim 20,000\ \mathrm{km\ s^{-1}}$ (e.g., \citeNP{Peterson1997}).
From the observed strength of UV emission lines, we know that the temperature
of the emitting plasma is on the order of a few times $10^4\ \mathrm{K}$
(e.g., \citeNP{Osterbrock1989}), insufficient to produce the observed line widths
via thermal (Doppler) broadening.  Instead, bulk motions of the BLR gases
appear to be responsible for the line broadening.

In AGNs, the BLR gases have apparently condensed into clouds, but the
medium surrounding Sgr A* does not share this property.  \citeN{FromerthMelia2001} 
have explored various scenarios for the AGN cloud formation based on the 
underlying principle that the source of plasma is ultimately that portion of 
the gas trapped by the central black hole from the interstellar medium.  Winds 
accreting onto a central black hole are subjected to several disturbances 
capable of producing shocks, including a Bondi-Hoyle flow, stellar wind-wind 
collisions, and turbulence.  Shocked gas is initially compressed and heated 
out of thermal equilibrium with the ambient radiation field; a cooling 
instability sets in as the gas is cooled via inverse-Compton and bremsstrahlung 
processes.  If the cooling time is less than the dynamical flow time through 
the shock region, the gas may clump to form the clouds responsible for broad 
line emission seen in many AGN spectra.  In the case of Sgr A*, this time 
differential does not appear to be sufficient for the cloud condensation to 
occur in the gravitationally trapped gas.  For AGNs, however, the preliminary 
calculations in this study suggest that clouds form readily. Their distribution 
agrees with the results of reverberation studies, in which it is seen that the 
central line peak (due to infalling gas at large radii) responds slower to continuum 
changes than the line wings, which originate in the faster moving, circularized 
clouds at smaller radii.  Very interestingly, it appears that the required cloud 
formation is one in which ambient gas surrounding the black hole (e.g., from 
stellar winds) is captured gravitationally and begins its infall with a (specific 
angular momentum) $\lambda$ representative of a flow produced by many wind-wind 
collisions and turbulence (see Fig.~\ref{figwinddistrib}) rather than a smooth 
Bondi-Hoyle bow shock. In this process, the gas eventually circularizes at 
$r_{circ} \approx 2 \lambda^2 r_s$ (see \S\ 4), but by that time all of the 
BLR clouds have been produced, since at that radius the gas presumably settles 
onto a planar disk.  As such, this picture is distinctly different from 
``conventional'' models in which the clouds are produced within a disk and are 
then accelerated outwards by such means as radiation pressure or magnetic 
stresses.

Finally, the radio emitting region surrounding the black hole in the Galactic
Center may be similar to what we see in the cores of more luminous AGNs.
In all classes of accreting black holes, a fraction of sources produce
flat-spectrum radio cores. In quasars they have been studied with 
VLBI and have been resolved into relativistic jets \cite{Zensus1997}. 
We have not yet seen a jet in Sgr A*, and its existence would be irrelevant
to Sgr A*'s spectrum if the emission is dominated by the accreting gas.
It is interesting to note that the spiral galaxy M81 has
a radio core with properties not unlike those of Sgr A*. This source 
has been observed with VLBI and was recently resolved into 
an extremely compact, though stretched out, structure
\cite{BietenholzBartelRupen2000}. It has a similar, unusually high
circular-to-linear polarization ratio as Sgr A*
\cite{BrunthalerBowerFalcke2001}. 
The extended emission in this
source on scales of $\sim 10-100$ mas is less than $6\%$ of the total
flux density.  The size and orientation of this structure are frequency
dependent, bending from $\sim40^\circ$ at 22 GHz to $\sim75^\circ$
at 2.3 GHz. These characteristics still leaves the question open as to
whether the dominant emitting region is inflowing or outflowing on
a compact scale. We may be seeing a combined core-jet emitter, in 
which the outflow contributes at least partially to the overall flux. 
In general, Sgr A*-like radio cores seem to be rather common in nearby
galaxies with a low level of nuclear activity (see, e.g.,
\citeNP{WrobelHeeschen1984,NagarFalckeWilson2000,FalckeNagarWilson2000}),
indicating that Sgr A* can tell us a great deal about the active
nuclei of other galaxies.

\section{CONCLUSIONS}
We have learned a great deal about the principal interactions within
the inner few parsecs at the Galactic Center, but as is often the case,
important questions arise with each uncovering of a new layer.  There
is no longer any doubt that a significant concentration of dark matter
occupies the region bounded by the inner $0.015$ pc.  This size is
sufficiently small that we can rule out distributions of stellar-sized objects,
such as neutron stars or brown dwarfs as the constituents.  Such a distribution
would need to be highly peaked in the center, and therefore considerably out
of equilibrium \cite{GenzelEckartOtt1997}.  Its lifetime would be of order
$10^7$ years, much smaller than the age of the galaxy \cite{Maoz1998}, 
leaving us to ponder why we are viewing this region at such a special time.  
That there is a massive point-like object in the middle is now hard to dispute.
It doesn't move relative to objects around it, and it has a spectrum like
no other in the Milky Way, though it shares many characteristics in common
with the cores of other nearby galaxies.  

One of the principal problems
now facing us is to understand how in fact Sgr A* produces its spectrum.
The Galactic Center is rich in gas, and some of it must be funneling into
the black hole.  Yet this process does not appear to be converting very
much kinetic and gravitational energy into radiation, making Sgr A*
extremely sub-Eddington.  This departure from
our naive expectations is forcing us to rethink the basic elements of 
accretion physics.  So theorists are now grappling with questions such
as (1) is the inflow advection dominated, carrying most of its energy
through the event horizon? (2) is the assumption of equipartition between
the magnetic field and the gas an over-simplification that leads to a great
overestimation of the magnetic field intensity, and hence of the synchrotron
emissivity? (3) does the plasma separate into two temperatures as it gets 
compressed and heated?  and (4) does the black hole and/or the infalling
plasma produce a jet at small radii that then dominates the emissivity from
this source?  Ongoing polarimetric observations at mm and sub-mm wavelengths
will greatly assist in this endeavor, providing the necessary constraints 
that are complementary to those implied by the spectroscopic measurements.  

Perhaps one of the most exciting developments in this program will be the
imaging of Sgr A*'s shadow against the backdrop of optically thin emitting
plasma at sub-mm wavelengths within the next $5$ to $10$ years. The appearance
of this shadow is a firm prediction of General Relativity, which mandates
a unique shape and size for the region where light bending and capture are
important.  There has never been such an opportunity to place the existence
of black holes on such a firm footing.  Galactic black hole binaries contain
compact objects that are too small, and the cores of other galaxies are simply
too far away.  Sgr A* at the Galactic Center has a size that is now on 
the verge of detectability with sub-mm VLBI.  This coming decade may finally
give us a view into one of the most important and intriguing predictions of
General Relativity. 

\vskip 0.2in
\hbox{\it Acknowledgments}\par
The authors are indebted to the vibrant Galactic Center community
for the many discussions and opportunities to learn about the latest
developments in this exciting field.  They are particularly grateful
to Don Backer, Geoff Bower, Dan Gezari, Mark Morris and Charles 
Townes for carefully reading
through the text and making helpful suggestions. This work was supported by a 
Sir Thomas Lyle Fellowship and a Miegunyah Fellowship for distinguished 
overseas visitors at the University of Melbourne, and by NASA grants 
NAG5-8239 and NAG5-9205 at the University of Arizona.

\bibliography{gcreview}
\bibliographystyle{apj}



\end{document}